%% file: main.tex
\documentclass[aps,pra,twocolumn,final,floatfix,superscriptaddress,10pt]{revtex4-2}

\usepackage[nolist]{acronym}
\usepackage{physics}
\usepackage{caption}
\usepackage{subcaption}
\usepackage{amsmath}
\usepackage{algorithm}
\usepackage{graphicx}  
\usepackage{xcolor}
\usepackage{tabularx}
\usepackage[normalem]{ulem}
\usepackage[utf8]{inputenc}
\usepackage{microtype}
\usepackage{todonotes}
\usepackage{algpseudocode}
\usepackage{mathtools}
\usepackage{relsize}
\usepackage{lipsum}
\usepackage{braket}

\usepackage{tikz}
\usetikzlibrary{
	matrix, 
	quantikz,
}

\definecolor{dlrgray}{RGB}{70,70,70}
\usepackage{mathtools}

\newcommand{\iu}{\mathrm{i}}

\definecolor{sbblue}{rgb}{0.2823529411764706, 0.47058823529411764, 0.8156862745098039}
\definecolor{sborange}{rgb}{0.9333333333333333, 0.5215686274509804, 0.2901960784313726}
\definecolor{sbgreen}{rgb}{0.41568627450980394, 0.8, 0.39215686274509803}
\definecolor{sbred}{rgb}{0.8392156862745098, 0.37254901960784315, 0.37254901960784315}
\definecolor{sbpurple}{rgb}{0.5843137254901961, 0.4235294117647059, 0.7058823529411765}
\definecolor{sbbrown}{rgb}{0.5490196078431373, 0.3803921568627451, 0.23529411764705882}
\definecolor{sbmagenta}{rgb}{0.8627450980392157, 0.49411764705882355, 0.7529411764705882}
\definecolor{sbgray}{rgb}{0.4745098039215686, 0.4745098039215686, 0.4745098039215686}
\definecolor{sbocca}{rgb}{0.8352941176470589, 0.7333333333333333, 0.403921568627451}
\definecolor{sblightblue}{rgb}{0.5098039215686274, 0.7764705882352941, 0.8862745098039215}

\begin{document}

	\title{Coherent and non-unitary errors in $ZZ$-generated gates}
	\author{T. Mueller} \email{Thorge.Mueller@dlr.de} 
	\affiliation{
German Aerospace Center (DLR), Institute for Software Technology, Department High-Performance Computing, 51147 Cologne, Germany} \affiliation{Theoretical Physics, Saarland University, 66123 Saarbr{\"u}cken, Germany}
	
	\author{T. Stollenwerk} 
	\affiliation{Institute for Quantum Computing Analytics (PGI 12),Forschungszentrum J{\"u}lich, 52425 J{\"u}lich, Germany}
	
	\author{D. Headley} 
	\affiliation{Theoretical Physics, Saarland University, 66123 Saarbr{\"u}cken, Germany}
	\affiliation{Institute for Quantum Computing Analytics (PGI 12),Forschungszentrum J{\"u}lich, 52425 J{\"u}lich, Germany}
	
	\author{M. Epping} 
	\affiliation{
German Aerospace Center (DLR), Institute for Software Technology, Department High-Performance Computing, 51147 Cologne, Germany}  
	
	\author{F. K. Wilhelm} 
	\affiliation{Theoretical Physics, Saarland University, 66123 Saarbr{\"u}cken, Germany}
	\affiliation{Institute for Quantum Computing Analytics (PGI 12),Forschungszentrum J{\"u}lich, 52425 J{\"u}lich, Germany}
	\date{November 2019}
	\date{\today}
	
	\begin{abstract}
		Variational algorithms such as the Quantum Approximate Optimization Algorithm have attracted attention due to their potential for solving problems using near-term quantum computers. The $ZZ$ interaction typically generates the primitive two-qubit gate in such algorithms applied for a time, typically a variational parameter, $\gamma$. Different compilation techniques exist with respect to the implementation of two-qubit gates. Due to the importance of the $ZZ$-gate, we present an error analysis comparing the continuous-angle controlled phase gate (\textsc{CP}) against the fixed angle controlled $Z$-gate (\textsc{CZ}).
		We analyze both techniques under the influence of coherent over-rotation and  depolarizing noise. 
		We show that \textsc{CP} and \textsc{CZ} compilation techniques achieve comparable $ZZ$-gate fidelities if the incoherent error is below $0.03 \, \%$ and the coherent error is below $0.8 \, \%$.
		Thus, we argue that for small coherent and incoherent error a non-parameterized two-qubit gate such as \textsc{CZ} in combination with virtual $Z$ decomposition for single-qubit gates could lead to a significant reduction in the calibration required and, therefore, a less error-prone quantum device. We show that above a coherent error of $0.04 \pi$ ($2 \, \%$), the \textsc{CZ} gate fidelity depends significantly on $\gamma$. 
	\end{abstract}
	
	\maketitle
	
	\section{Introduction}
	Quantum computers give rise to a new class of algorithms that are not efficiently simulable on classical computers due to the exponential scaling of classical memory required. Quantum algorithms such as Shor's \cite{Shor} and Grover's \cite{Grover_first} are interesting candidates in the anticipated era of fault-tolerant quantum computing, providing guaranteed asymptotic speedups versus the best-known classical counterparts. In an era of fault-tolerant quantum computers, device error rates fall sufficiently below threshold values, and therefore quantum error correction codes can be utilized.
	Improvements at all levels are needed to pass this fault-tolerance threshold, from the materials used to fabricate qubits to the on-device layout of physical qubits and the high-order quantum logic performed. 
	Contemporary superconductor-based hardware has attained fidelities for a two-qubit gate in excess of 99 \% \cite{supremacy} and, as such, current hardware fidelities come close to the error-threshold \cite{Barends_Martinis}. Because of widely spread and popularity, we focus on superconducting platforms and their typical available two-qubit gates.  
	Error sources of two-qubit gates must be analyzed to improve in the coherent and incoherent error regime.
	In recent years variational quantum algorithms \cite{farhi2014quantum,Peruzzo2014} have gained more attention due to their potential to be useful in the NISQ era. Fault-tolerant algorithms, while providing more concrete performance guarantees, are of limited interest as it is unlikely that hardware will advance to enable their use in the near future \cite{FTQC_limits}. Variational NISQ algorithms provide modest hope that with a relatively small number of physical qubits ($\approx 100$) in the absence of error correction, errors, especially those which are coherent, could be mitigated. 
    This work is structured as follows: First, we give a brief overview of the possible compilation strategies for ZZ-gate into \textsc{CZ}, i\textsc{SWAP}, and \textsc{CP} gates. Second, we introduce error channels included in our study for coherent and incoherent errors. Third and finally, we present numerical and analytical results using error channels for \textsc{CZ} and \textsc{CP} decompositions under differing error conditions. We conclude which decomposition strategy is likely to provide greater fidelity.

	\section{QAOA}
	An algorithm of particular interest is the Quantum Approximate Optimization Algorithm (QAOA) \cite{farhi2014quantum}. QAOA is a heuristic algorithm aiming to find high-quality solutions to combinatorial optimization problems. It is among the most promising candidates to show quantum supremacy \cite{supremacy_QAOA} in the near future.
	A multitude of studies, both numerical \cite{qaoa_analytic_sim} and analytical \cite{qaoa_analytic,qaoa_analytic_II}, have been performed for QAOA in the ideal zero-error case. Such studies show that while QAOA is universal---any unitary transformation may be expressed in QAOA sequences with driver and problem Hamiltonians---a minimum circuit depth will typically be required to reach the optimal solution of an encoded problem. One important use case of the QAOA algorithm is the approximate solution of MAX-CUT problems, that is, to find low energy states of the problem Hamiltonian
	\begin{equation}
	    H_{\mathrm{P}} = \sum_{(i, j)\in E} Z_i Z_j \,,
	\end{equation}
    where $E$ are the edges in a specified problem graph. QAOA consists of layers of alternating Hamiltonians. The first, the driver Hamiltonian, typically takes the form of single-qubit $X$ rotation gates applied to each qubit. These gates, often described as the transverse field operators in quantum annealing literature\cite{transverse_field}, are responsible for inducing transitions between computational basis states. 
    and therefore solution states. The QAOA ansatz state is constructed with $p$ pairs of alternating unitaries, determined by the problem and transverse field Hamiltonians,
	\begin{equation}
		\ket{\boldsymbol{\gamma,\, \beta}}_p = U_{\mathrm{M}}(\beta_p) U_{\mathrm{P}}(\gamma_p)\dots U_{\mathrm{M}}(\beta_1)U_{\mathrm{P}}(\gamma_1) \ket{s} \,,
	\end{equation}
	where ${\boldsymbol{\gamma}, \boldsymbol{\beta}}$ are classical parameters to be optimized and 
		\begin{equation}\label{eqn:QAOA}
		U_{\mathrm{M}}(\beta) = \exp(-i \beta \sum_i X_i), \quad
		U_{\textrm{P}}(\gamma) = \exp(-\iu \gamma H_P) \, ,
	\end{equation}
	and $\ket{s} = \textrm{H}^{\otimes n} \ket{0}^{\otimes n}$.
 While most previous works have not considered the effects of imperfect engineering and interactions with the environment, coherent or otherwise, on the performance of QAOA, some authors have analyzed the effects of depolarizing error models on QAOA performance \cite{streif2020quantum,Marshall_2020}. No study, however, has considered the influence of errors for different gate decompositions, as is the subject of this work. Two-qubit $ZZ$-gates encode the problem Hamiltonian of interest and evolve the phase of computational basis states with a dependence on an objective function to be optimized.
	\section{Sources of Noise in QAOA}
    We have two different types of error sources to distinguish. The first one introduces coherent errors. This error represents coherent unitary over-rotations in the system. The second type of error source causes an unwanted non-unitary evolution of the system and introduces incoherent errors. They are typically due to interactions with an environment.
    Errors stemming from the single-qubit $X$ rotations are of little interest as such gates are typically executed one or more orders of magnitude faster and more precisely than interacting gates regardless of the platform \cite{charge_noise,Tosi2017,Barends_Martinis}. As interactions with the environment typically induce noise proportional to the duration of the interaction, such gates result in negligible errors.
	The $ZZ$-gate
	\begin{equation}
 R_{ZZ}(\gamma)=
	    \left(\begin{array}{cccc} 
	    1 & 0 & 0 & 0\\
	    0 & e^{  i  \textcolor{black}{\gamma}} & 0 & 0\\ 
	    0 & 0 & e^{  i  \textcolor{black}{\gamma}} & 0\\ 
	    0 & 0 & 0 & 1 
	    \end{array}\right) 
	\end{equation}
	is the two-qubit gate used in the algorithm and is the focus of our investigation of QAOA under the influence of noise. Two-qubit gates are more error-prone due to their  more complicated design than single-qubit gates \cite{supremacy}. 
	Some two-qubit gates are constructed by exploiting direct qubit-qubit interactions. Some utilize intermediate ancillary components as in a tunable coupler design \cite{tubale_coupler}, in which one inserts an ancilla qubit leading to additional routes for the environment to interact with the system and potentially greater error.
	Furthermore, such gates typically increase execution times \cite{speed_ratio}, leading to a higher probability of incoherent errors. 
     Different compiling strategies exist for the $ZZ$-induced gate depending on which two-qubit gates are natively available on a device, with different groups using different approaches. The Google-developed hardware platform supports a novel \textit{fsim} gate \cite{google_fsim}, which is an $XY$-gate with a phase shift on the last diagonal element. Devices manufactured by Rigetti support an extended family of $XY$-gates \cite{rigetti} using a single calibrated two-qubit gate \cite{rigetti}.
     The Wallraff group at ETH Z{ü}rich provides the parametric \textsc{CP} gate \cite{Wallraff_cp_cz}, and the IBM platform uses a \textsc{CNOT} gate. 
	We investigate the influence on the performance of coherent over-rotation and depolarizing errors in connection with different compilation platforms. We focus primarily on the decomposition of the $ZZ$-gate with parametric \textsc{CP} versus fixed-angle \textsc{CZ} gate as these gates require different numbers of two-qubit gates to simulate a $ZZ$-interaction. While the variable angle \textsc{CZ} gate decomposition needs two fixed \textsc{CZ} gates, the parametric \textsc{CP} gate requires one.
     \section{Implementations of the $ZZ$-gate}\label{sec:error_impl}
	Depending on the native two-qubit gate in a specific hardware platform, the decomposition of the $ZZ$-gate takes different forms. In this section, we describe the decompositions into i\textsc{SWAP}, \textsc{CZ}, and \textsc{CP} gates, which are typically available on superconducting devices. In contrast to the \textsc{CP} and \textsc{CZ}, the decomposition into i\textsc{SWAP} is less well-known. First, we investigate $R_{ZZ}(\gamma)$ into \textsc{CP} decomposition 
	\begin{equation}\label{eqn:cphase}
	 R_{ZZ}(\gamma) = 
		\begin{quantikz}[ampersand replacement=\&]
			i \& \&  \gate{R_Z( \textcolor{black}{\gamma})}  \& \gate[wires=2][1cm]{\text{CP}(-2 \gamma)} \& \qw \\
			j \& \&  \gate{R_Z( \textcolor{black}{\gamma})}  \&   \& \qw
		\end{quantikz} \, ,
	\end{equation}
         \begin{equation}
         \text{with} \quad
         \text{CP}(\gamma) = 
		\begin{pmatrix}
			1 & 0 & 0 & 0\\
			0 & 1 & 0 & 0\\
			0 & 0 & 1 & 0\\
			0 & 0 & 0 & e^{ i \gamma}
		\end{pmatrix}. \\
          \end{equation}
    The single-qubit Z-rotation gate $R_{Z}(\gamma)$ is defined by 
    \begin{equation}
        R_{Z}(\gamma)=
	    \left(\begin{array}{cc} 
	    1 & 0 \\
	    0 & e^{  i  \textcolor{black}{\gamma}}\\
	    \end{array}\right). 
    \end{equation} 
     The circuit diagram of $R_{ZZ}(\gamma)$ in \textsc{CP} decomposition can also be written as $R_{ZZ}(\gamma) = \text{CP}(- 2\gamma) R_{Z_{1}}(\gamma) \, R_{Z_{2}}(\gamma)$. The Wallraff group introduced this decomposition \cite{Wallraff_cp_cz} consisting of two single-qubit $Z$-rotations depending on the variational parameter $\gamma$ and the controlled \textsc{CP} gate, which also depends on $\gamma$. 
     The parameterized two-qubit gate could lead to an overhead in the calibration compared to a fixed-parameter two-qubit gate like the \textsc{CZ}. The decomposition into \textsc{CZ} is according to
     \begin{equation} \label{eq:ps_layer}
	R_{ZZ}(\gamma) = 
		\begin{quantikz}[row sep=0.5em, column sep=0.4em]
			i\; & \qw & \qw      & \ctrl{1}    & \qw      & \qw                  & \qw      & \ctrl{1}    & \qw      & \qw \\
			j\; & \qw & \gate{H} & \control{}  & \gate{H} & \gate{R_Z( \gamma)} & \gate{H} & \control{}  & \gate{H} & \qw
		\end{quantikz}
		\,.
	\end{equation}
     The decomposition comprises five single-qubit gates and two fixed two-qubit \textsc{cz} gates. The rotation angle $\gamma$ is contained in only one single-qubit $Z$-rotation. 
    The decomposition into i\textsc{SWAP} requires two two-qubit gates \cite{cooks} because i\textsc{SWAP} does not belong to the same equivalence class as the $ZZ$-gate. Furthermore, we demand the i\textsc{SWAP} decomposition to have the following single-qubit rotations: Hadamard and Pauli-Z rotation gate. We further know that the rotation gate containing $\gamma$ will be sandwiched by the two i\textsc{SWAP} gates. With the help of work concerning $XY$-based interactions \cite{regensburg} it is straightforward to show that the compilation from a $ZZ$-gate into i\textsc{SWAP} reads
	\begin{widetext}
		\centering
		\begin{equation}
			\label{eq:ps_layer}
			R_{ZZ}(\gamma) = 
			\begin{quantikz}[row sep=0.5em, column sep=0.5em]
				i\; & \qw       & \gate{R_Z(\pi / 2)} & \gate[wires=2][1.5cm]{\mathrm{iSWAP}}  & \gate{H}  & \gate{R_Z( \gamma)} & \gate{H}  & \gate[wires=2][1.5cm]{\mathrm{iSWAP}}& \gate{R_Z( \pi / 2)} & \qw      & \qw \\
				j\; &  \gate{H} & \gate{R_Z(- \pi / 2)} & \qw                                    & \qw       & \qw                & \qw       & \qw                                  & \gate{R_Z(- \pi / 2)} & \gate{H} & \qw
			\end{quantikz}
			\,.
		\end{equation}
	\end{widetext}
	The i\textsc{SWAP} gate belongs to the $XY$ family ($XY(\theta = 90)$).
 Accompanying the two i\textsc{SWAP} gates in the decomposition are 8 single-qubit gates and a Z-rotation which confers the rotation angle $\gamma$. Another hardware context in which such gates might be available is that of trapped ion-based quantum computers. These machines natively use a Mølmer–Sørensen gate
	\begin{equation}
	    MS(\gamma) = \exp(i\gamma\sum_{l,k} X_{l} X_{k})
	\end{equation} 
	 and only need single-qubit rotations to achieve the $ZZ$-gate \cite{MS}. This could also lead to an efficient QAOA compilation. However, this hardware design differs too much from the gate level with superconducting circuits to include it in our comparison. The error sources are also different for these two architectures.
	\section{Depolarizing and coherent errors in \textsc{CP} and \textsc{CZ}}
	Now we introduce the error channels we apply to our $ZZ$-gate operations. On the one hand, we want to simulate a coherent error that we incorporate by over-rotation angles $\theta,\zeta$ in controlled-phase gates $\textrm{CP}(\theta),\textrm{CP}(\zeta)$. On the other hand, we apply a two-qubit symmetric depolarizing error channel to our two-qubit gates: 
	\begin{equation}
	\mathcal{E}(\rho) = \dfrac{p \mathit{I}_{4}}{4}+(1-p) \rho,
	\end{equation} 
	with $\mathit{I}_{4}$ being the identity matrix for the two-qubit Hilbert space and $p$ the probability that the density matrix $\rho$ will end up in a total mixed state. We assume a Markovian, Pauli-error channel. For simplicity of the simulations, we use the Kraus Operator representation 
     \begin{equation}\label{eqn:krauss}
		\mathcal{E}(\rho) = \mathlarger{\sum_{i}^{16}} m_{i} K_{i} \rho K_{i}^{\dagger} \quad \text{with} \quad \mathlarger{\sum_{i}^{16}} m_{i} K_{i}K_{i}^{\dagger} = \mathit{I}_{4}
	\end{equation}
   and $K_{i}=\omega_{\lfloor \frac{i}{4} \rfloor} \otimes \omega_{i \bmod 4}$ with $\omega = (\mathit{I}_{4},X,Y,Z)$. We define $m_{1}=\sqrt{1-(15/16)p}$ and otherwise $m_{i}=\sqrt{p/16}$. $X,Y,Z$ are the Pauli matrices. This is the typically chosen Kraus decomposition for the depolarizing error, which is not unique. 
   We set the incoherent error to a max value of $p=1 \, \%$  because the gate fidelity declines below $99 \, \%$ for $p>1 \, \%$. Below $ 99 \, \%$, the gate is not suitable for a fault-tolerant quantum computer \cite{threshold_2}.
	The \textsc{CP} gate's coherent error we take into account by adding the error angle $\theta$ to $-2 \gamma \rightarrow - 2 \gamma + \theta$ leading to the error-prone unitary operation
		\begin{equation}\label{eqn:coherent_cp}
		U_{cp}^{co} = R_{Z_{2}}(\gamma) \, R_{Z_{1}}(\gamma) \, \text{CP}(- 2 \gamma + \theta).
	\end{equation}
	The coherent error for the \textsc{CZ} gate decomposition we are taken into account by adding a \textsc{CP} with the coherent error phases $\theta,\zeta$ after the first and the second CZ gate, respectively, leading to the coherent error unitary
		\begin{equation}\label{eqn:coherent_cz}
		U_{cz}^{co} = \text{H}_{2} \, \text{CZ} \, \text{CP}(\theta) \, \text{H}_{2} R_{Z_{2}}(\gamma) \, \text{H}_{2} \, \text{CZ} \, \text{CP}(\zeta) \, \text{H}_{2}.
	\end{equation}
	The subscripts of $U_{cp}^{co}$ and $U_{cz}^{co}$ describe the gate decomposition and the superscript, the kind of error. For coherent errors, the superscript is $co$, and for decoherent errors $de$. 
	For the error range in the coherent case, we decide to go from 0 to $0.06 \pi$ ($ \approx 3 \, \% $). This is a realistic error range for superconducting platforms. The over-rotation angles $\theta,\zeta$ are picked independently from a Gaussian function with zero mean and standard deviation $\sigma( \theta)$ or $\sigma( \zeta)$.\\
    The gate fidelity is defined as the average of the integral over all state fidelities
	\begin{equation}
	F(\rho,\rho^{\prime}) = \left(\text{Tr} \left(\sqrt{\sqrt{\rho} \rho^{\prime} \sqrt{\rho}} \right) \right)^{2}.
	\label{eqn:state_fidelity}
    \end{equation}
     $\rho^{\prime}$ and $\rho$ are the output density matrices after the gate operation with and without error. Cabera et al. \cite{CABRERA,Bowdrey} showed that the integral for the gate fidelities reduces to a sum over 16 initial input states $\ket{\psi_{a}}_{1} \ket{\psi_{b}}_{2}$  (a,b = 1,..,4)) with
	\begin{equation}
	    \begin{split}
	         &\ket{\psi_{1}} = \ket{0}, \ket{\psi_{2}} = \ket{1} , \\
	 &\ket{\psi_{3}} =\frac{1}{ \sqrt{2}}  ( i \ket{1}+ \ket{0} ) , \ket{\psi_{4}} =\frac{1}{ \sqrt{2}} (\ket{1}+ \ket{0} ).
	    \end{split}
	\end{equation}
    Furthermore, if we consider that the output state after applying the error-free gate operation is always pure, we deduce from eq. (\ref{eqn:state_fidelity}) the gate fidelity
	\begin{equation}\label{eqn:gate_fidelity}
		\mathcal{F} = \frac{1}{16}\sum_{j=1}^{16} 
		\Braket{\psi_{j}|U^{^{\dagger}}\rho_{j}^{\prime}U|\psi_{j}}.
	\end{equation}
	$\rho_{j}^{\prime}$ is the state $\psi_{j}$ after applying the decomposition of \textsc{CZ} or \textsc{CP} with the error. In contrast to $U$, which describes the gate operation without error. We first apply the error channels separately and then together to analyze the different effects of both errors. The iSWAP decomposition consists of two fixed two-qubit gates like CZ. We are not expecting a different gate fidelity compared to CZ decomposition. CZ and iSWAP decomposition only differs in single-qubit gates, and the error channels do not affect single-qubit gates. Only the state fidelities will change. For this reason, we are not investing in numerical and analytical studies for iSWAP decomposition and refer to CZ decomposition instead.

	\section{Numerical studies and analytical results}
\begin{figure}
		\includegraphics[scale =1.0]{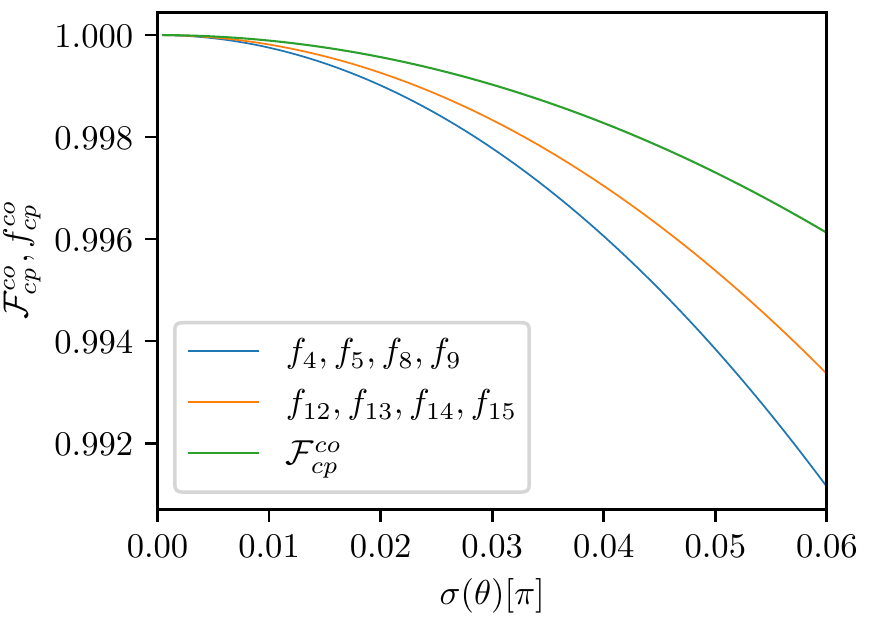}
		\caption{Gate $\mathcal{F}^{co}_{cp}$ and state $f^{co}_{j}$ fidelities (y-axis) plotted against the standard deviation $\sigma(\theta)$ for the Gaussian coherent error (x-axis) for $\textsc{CP}$ decomposition. The graph shows the gate fidelity and the split into 16 state fidelities. The solid green line represents the gate fidelity $\mathcal{F}_{cp}^{co}$, and the orange and blue line represent the state fidelities. The error-unaffected state fidelities are not shown.}\label{fig:coherentcp}
	\end{figure}

	In this section, we apply the coherent and incoherent error channels to the ZZ-gate decomposition we have introduced before. We compare how the gate fidelities change. We analyze the gate fidelity for pure coherent over-rotation, see eq. ~\eqref{eqn:coherent_cp} and ~\eqref{eqn:coherent_cz}. The quantum fault-tolerance theorem makes no statement about the error threshold specifically for coherent errors but for incoherent errors. 
    The limit for the worst case error is lower in the coherent case due to more efficiency for existing surface codes \cite{Performance_QEC_coherentErrors_II, Performance_QEC_coherentErrors_III,Performance_QEC_coherentErrors}. 
	We can rewrite eq. (\ref{eqn:gate_fidelity}) to
	\begin{equation} \label{eqn:gate_fidelity_coherent}
	    	\mathcal{F}^{co} = \frac{1}{16}\sum_{j=1}^{16} |\braket{\psi_{j} | U^{\dagger}U^{co} | \psi_{j}}|^{2}, 
	\end{equation}
	with $U^{\dagger}$ the error-free gate operation and $U^{co}$ the coherent error operation. If we now decompose into \textsc{CP} gate, we derive the gate fidelity  
     \begin{equation}
	\begin{split}
		\mathcal{F}_{cp}^{co} &= \frac{1}{16}\sum_{j=1}^{16} |\braket{\psi_{j} | R_{ZZ}^{\dagger}(\gamma) \, U^{co}_{cp} | \psi_{j}}|^{2} \\
		&= \frac{1}{16} \sum_{j=1}^{16} |\bra{\psi_{j}} R_{Z_{1}}^{\dagger}(\gamma) \, R_{Z_{2}} ^{\dagger}(\gamma) \, \text{CP}^{\dagger}(\gamma) \\
		& \times R_{Z_{1}}(\gamma) \, R_{Z_{2}} \, \text{CP}(\gamma+\theta)  \ket{\psi_{j}}|^{2}\\
		&= \frac{1}{16} \sum_{j=1}^{16} |\braket{\psi_{j} | \text{CP}(\theta)| \psi_{j}}|^{2}.
	\end{split}	
\end{equation} 
	The gate fidelity $\mathcal{F}_{cp}^{co}$ is independent of the rotation angle $\gamma$. This is because the error $\textrm{CP}(\theta)$ as well as the error-free gate Operation $U_{cp}$ are diagonal. The gate operations can thus commute through and cancel out. We can now calculate the 16 state fidelities. We achieve three different state fidelity equations for all 16 input states: $f_{1,2,3,6,7,10,11,16} = 1,f_{12,13,14,15}=(2+2 \, \text{cos}( \theta))/4,f_{4,5,8,9}=(10+6 \, \text{cos}( \theta))/16$. By averaging over all three types for their proportion, we derive 
      \begin{equation}
      \label{eqn:cp_coherent}
	    \mathcal{F}_{cp}^{co} = \frac{1}{32}  ( 25 + 7 \,  \cos( \theta) ).
	\end{equation}
   Fig. \ref{fig:coherentcp} shows the results for eq. (\ref{eqn:cp_coherent}). 
   The decomposition into $\textsc{CZ}$ gates is not diagonal, and therefore, the error channel $\textrm{CP}( \theta )$ cannot commute through and cancel out. If we insert the error-free and erroneous gate operation, see eq. (\ref{eqn:coherent_cz}), for the $\textsc{CZ}$ decomposition into eq. (\ref{eqn:gate_fidelity_coherent}) we derive
   \begin{equation}
	 \begin{split}
   		\mathcal{F}_{cz}^{co} &= \frac{1}{16}\sum_{j=1}^{16} |\braket{\psi_{j} | (R_{ZZ}^{\dagger}(\gamma)U_{cz} | \psi_{j}}|^{2} \\
     &= \frac{1}{16}\sum_{j=1}^{16} |\braket{\psi_{j} | (R_{ZZ}^{\dagger}(\gamma)  \,  \text{H}_{2} \, \text{CZ}  \, R_{X_{2}}(\gamma) \, \text{CZ} \, \text{H}_{2}|\psi_{j}}|^{2}.
	    \end{split}
	\end{equation}
    The closed solution for the gate fidelity$\mathcal{F}_{cz}^{co}$ exceeds the space constraint. Therefore, we assume small angles $\theta,\zeta$. For the gate fidelity of $\textsc{CZ}$, we calculate the 16 state fidelities up to the second order in $\theta,\zeta$. 
    If we sum up all state fidelities and make the small angle approximation for $\zeta$ and $\theta$, we derive the following equation for the coherent error in $\textsc{CZ}$ decomposition
	\begin{equation}\label{eqn:gate_fidelity_cz}
   	\begin{split}
   		\mathcal{F}_{cz}^{co} &= \frac{1}{16}\sum_{j=1}^{16} |\braket{\psi_{j} | U_{cz}^{\dagger}U_{cz}^{co} | \psi_{j}}|^{2}\\
   		&=  1 - 0.12 \, \theta^{2} - 0.13 \, \zeta^{2} - 0.05 \, \theta \zeta \\ &- 0.02  \, \theta \zeta \text{sin}{\left( \gamma \right)} - 0.19 \, \theta \zeta \text{cos}{\left(\gamma \right)} 
   		 - 0.02 \, \zeta^{2} \text{sin}{\left( \gamma \right)} \\
   		 &+ 0.02  \, \zeta^{2} \text{cos}{\left( \gamma \right)} +\mathcal{O}(h),
   	\end{split}
   \end{equation}
   with $h$ being all terms depending on $\zeta,\theta$ up to order three.
	\begin{figure}[h!]
		\centering
		\includegraphics[scale =1.0]{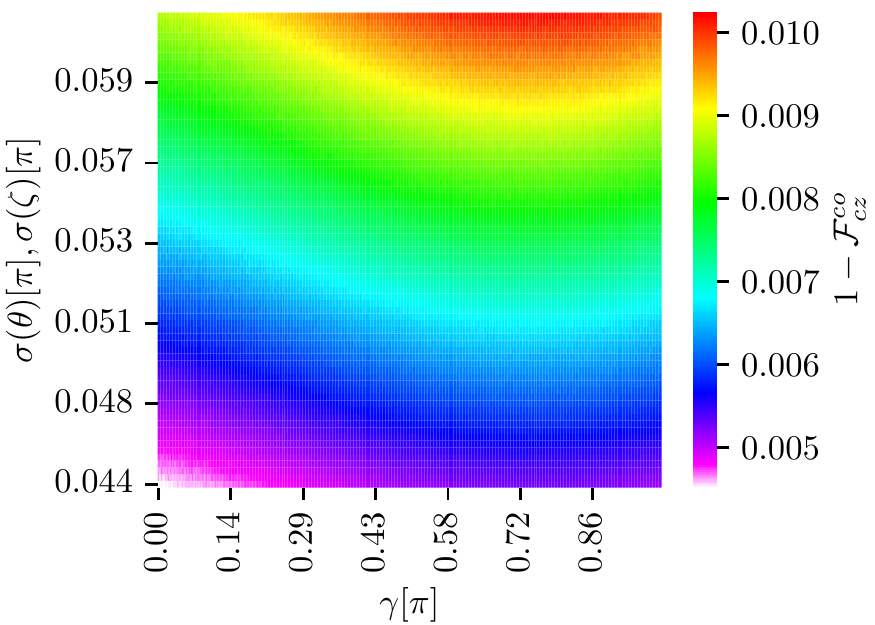}
		\caption{Gate infidelity 1-$\mathcal{F}_{cz}^{co}$ (colorbar) for the Gaussian coherent error with standard deviations $\sigma(\theta),\sigma(\zeta)$ (y-axis) and the rotation angle $\gamma$ (x-axis). The colorbar indicates the infidelity: a low value in the red regime(top-right) corresponds to a low fidelity of $99 \, \%$. The magenta regime's high value(bottom-left) relates to a high gate fidelity of $99.5 \, \%$. We average over 1000 repetitions per angle and error.}\label{fig:coherentcz}
	\end{figure}
	In contrast to $\mathcal{F}_{cp}^{co}$, $\mathcal{F}_{cz}^{co}$ 
	shows a weak dependency on the rotation angle $\gamma$. 
    Thus $\mathcal{F}_{cz}^{co}$ consists of small coherent over-rotation angles $\theta,\zeta$ coupled to $\gamma$ with $\sin( \gamma) \theta$ and $\cos( \gamma) \theta$.
    The numerical results for the \textsc{CZ} decomposition are shown in fig. \ref{fig:coherentcz}. In fact, for small $\sigma(\zeta),\sigma(\theta) < 0.04 \pi$, the dependency on $\gamma$ is weaker, and we recover the quadratic law from small error approximation of the cosine like for the \textsc{CP} decomposition. \\
    The minimum of the gate fidelity is at $\gamma = 0.72 \pi$ depending on $\sigma(\zeta),\sigma(\theta)$. For example, at this minimum, we attain a gate infidelity of $0.75 \, \%$ at a standard deviation $\sim$ $\sigma(\zeta),\sigma(\theta) = 0.054 \pi$ ($2.7 \, \%$). The same infidelity we derive at $\gamma = 0$ at a standard deviation $\sigma(\zeta),\sigma(\theta) = 0.0585 \pi $ ( $2.9 \, \%$). Conversely, if we fix $\zeta(\zeta),\sigma(\theta) = 0.056 \pi$, the gate infidelity is approximately $0.7 \, \%$ at a rotation angle of zero. The infidelity is $0.85 \, \%$ for the same standard deviation at the minimum of $\gamma = 0.72\pi$. Consequently, the small rotations angles $\gamma \approx 0$ and the larger ones  $\gamma \approx \pi$ are more error robust against coherent noise than close to  $\gamma \approx 0.72 \pi$. The gate fidelity difference in dependence of $\gamma$ is negligible for $\zeta(\zeta),\sigma(\theta) < 0.04 \pi$. The gate fidelity threshold of $99 \, \%$ is reached for a standard deviation $\sigma(\zeta),\sigma(\theta) > 0.06 \pi$. Further, we can simplify eq. (\ref{eqn:gate_fidelity_cz}) to 
	\begin{equation}\label{eqn:errors_analytic}
	 \mathcal{F}_{cz}^{co} = 1 -\theta^2 (0.3 + 0.04 \text{sin}(\gamma) + 0.17 \text{cos}(\gamma))+\mathcal{O}(\theta^{3}),
	\end{equation}
	if we assume $\zeta=\theta$. We can use this equation for potential extreme errors in one single $ZZ$-gate where both \textsc{CZ} gates face the highest possible error from the Gaussian error distribution.
	\begin{figure}[h!] 
	   \centering
		\includegraphics[scale =1.0]{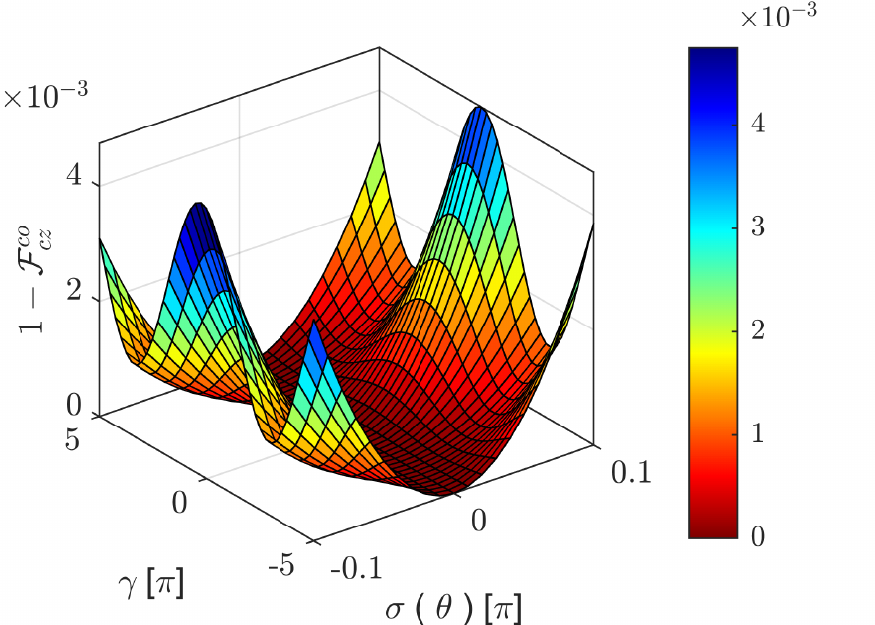}
		\caption{plot of eq. (\ref{eqn:errors_analytic}) for a standard deviation $\sigma ( \theta ) \in [ -0.1 \pi ,0.1\pi]$ and $\gamma \in [-1.5 \pi,1.5 \pi]$.}
		\label{fig:analytic_plot}
	\end{figure}
    Fig. \ref{fig:analytic_plot} allows us to estimate the gate infidelity for high and low standard deviations $\sigma(\theta)$ depending on the rotation angle $\gamma$.
    As we increase the standard deviation $\sigma(\theta)$, the variation in the gate fidelity through the rotation angle $\gamma$ increases.We plotted here the analytic function and not averaging over a randomly picked Gaussian distribution for $\sigma(\theta)$ as in fig. \ref{fig:coherentcz}. In this case, we see a shift in the numerical results in fig. \ref{fig:coherentcz}. Here we see the opposite side; gate fidelities are higher for small rotation angles $\gamma$ than for high rotation angles. But the overall trend in fig. \ref{fig:analytic_plot} is the same as for fig \ref{fig:coherentcz}.
     We now turn  to the case of a pure depolarizing error for \textsc{CP},
      \begin{equation}
	  F_{cp}^{de} = \frac{1}{16}\sum_{j=1}^{16} \bra{\psi_{j}} {U^{cp}}^{\dagger} \mathcal{E} (U^{cp}\ket{\psi_{j}} \bra{\psi_{j}}{U^{cp}}^{\dagger}) U^{cp}  \ket{\psi_{j}},
	  \end{equation}
	  and for \textsc{CZ} decomposition,
	  \begin{equation}
	  	F_{cz}^{de} = \frac{1}{16} \sum_{j=1}^{16} \bra{\psi_{j}} {U^{cz}}^{\dagger} \mathcal{E} ( \mathcal{E} (U^{cz}\ket{\psi_{j}} \bra{\psi_{j}}{U^{cz}}^{\dagger})) U^{cz}  \ket{\psi_{j}}.
	\end{equation}
	\begin{figure}[h!]
		\centering
		\includegraphics[scale =1.0]{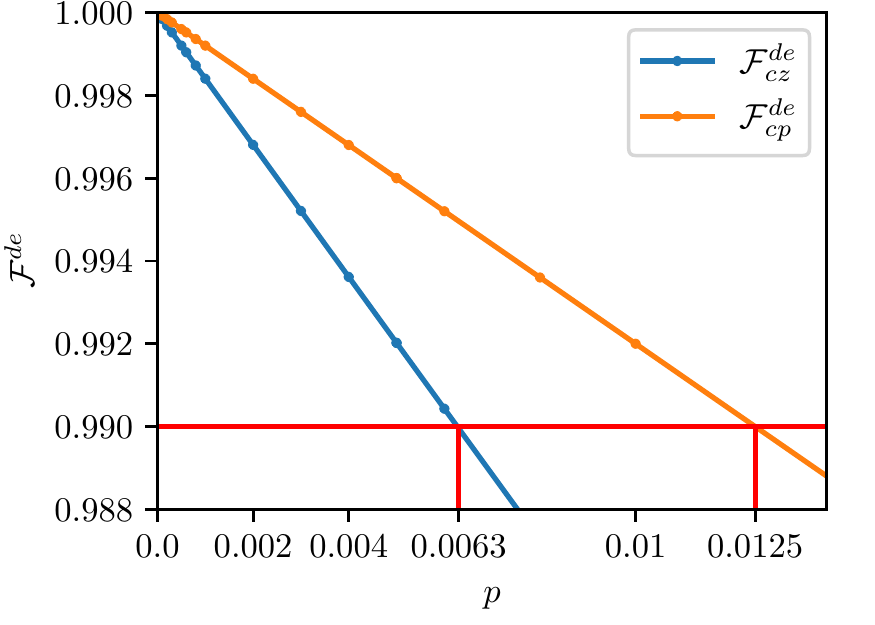}
		\caption{Gate fidelities plotted against depolarizing error for \textsc{CP} (orange) and \textsc{CZ} (blue) decomposition. The red line indicates the $99 \, \%$ gate fidelity.}\label{fig:dpl}
	\end{figure}
	Fig. \ref{fig:dpl} shows the gate fidelity behaviour under the depolarizing error for \textsc{CP} and CZ, $\mathcal{F}_{cp}^{de}$ and $\mathcal{F}_{cz}^{de}$, respectively. To accomplish a fully error-corrected quantum computer, the available two-qubit gate's fidelity must be well above the limit of $99 \, \%$  \cite{threshold_2,threshold_3}. The exact threshold value is an ongoing discussion.
    For specific connectivity graphs, this threshold could also be lower than $99 \, \%$ \cite{threshold_1}. $\mathcal{F}_{cz}^{de}$ drops below $99 \, \%$ gate fidelity for $p > 0.63 \, \%$. In contrast $\mathcal{F}_{cp}^{de}$ is more error robust. The threshold value of $99 \, \%$ is reached for $p=1.25 \, \%$.
    The linear behavior of $\mathcal{F}_{cp}^{de}=1-0.8 p$ follows directly from the definition of the symmetric two-qubit depolarizing error channel. There is no dependency on the rotation angle $\gamma$. The gate fidelity for $\mathcal{F}_{cp}^{de}=1-1.54 p$ is deduced by applying the error channel twice. After utilizing the channel a second time to the density matrix, the equation $\mathcal{F}_{cz}^{de}=1-(3 / 2)p+(3 / 4)  p^{2}$ is derived. For small error $p$, the linear scaling will be achieved. By applying the incoherent error channel twice, the limit for depolarizing error before crossing the $99 \, \%$ line is approximately doubled. In both cases, the single-state fidelities for all 16 states are equal.\\
	Next, we apply both error models simultaneously to both $ZZ$-gate decompositions.
	\begin{figure}[h!]
		\centering
		\includegraphics[scale =1.0]{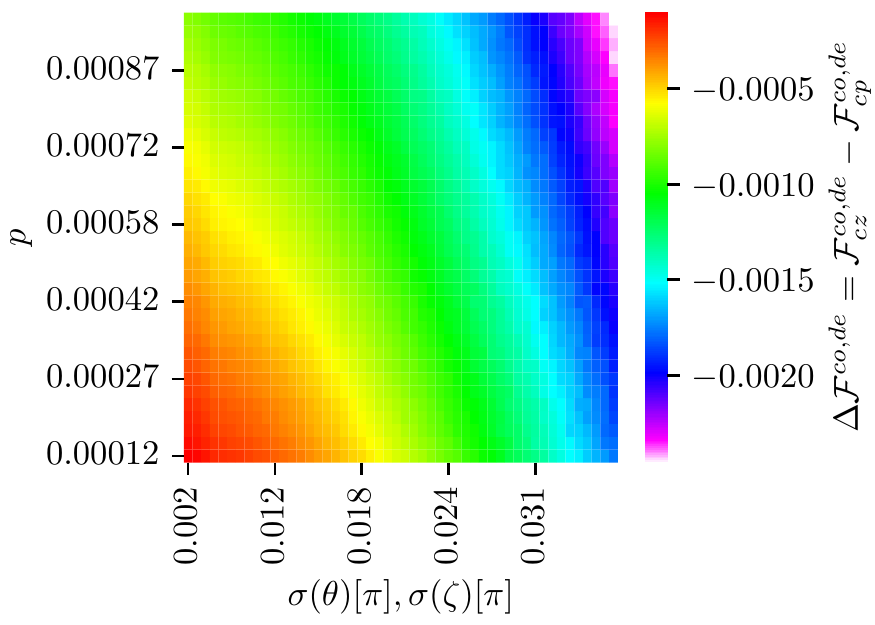}
		\caption{Plotting the fidelity difference $\mathcal{F}_{cp} - \mathcal{F}_{cz}$ in dependency on the coherent error's standard deviations $\sigma(\zeta),\sigma(\theta)$ (x-axis) and the depolarizing error $P$ (y-axis). magenta corresponds to a high advantage of $F_{cp}$ over $F_{cz}$ whereas red correspond to comparable gate fidelity of both decompositions.}\label{fig:de_coherentczcp}
	\end{figure}
	Fig. \ref{fig:de_coherentczcp} shows the difference in gate fidelities $\Delta\mathcal{F}^{de,co}$ between $\mathcal{F}_{cp}^{de,co}$,$\mathcal{F}_{cz}^{de,co}$ for coherent and depolarizing error. For fig. \ref{fig:de_coherentczcp}, we set the rotation angle to a small value of $\gamma = 0.01 \pi$ to achieve the greatest $F_{cz}^{co}$. The plot shows gate fidelities above $99 \, \%$.
	If $\sigma(\zeta),\sigma(\theta) < 0.016$ ($0.8 \, \%$) and $p < 0.032 \%$ the difference in the fidelities between both decompositions is $\Delta \mathcal{F}^{de,co} \approx 0.02 \%$. For standard deviations $\sigma(\zeta),\sigma(\theta) < 0.016$ ( $0.8 \, \%$), $p < 0.032 \%$ gate fidelities are of order $\mathcal{F}_{cz}^{de,co},\mathcal{F}_{cp}^{de,co} \approx 99.8 \%$. As a result, if we could suppress the incoherent error and allow for small coherent over-rotations, there is no advantage from \textsc{CP} over \textsc{CZ} decomposition. Of course, the circuit depth would double in the case of the \textsc{CZ} decomposition technique and could exceed coherence time. On the other hand, the pulse calibration for different angles for the parametric \textsc{CP} gate would also lead to another error source. 
    As hardware platforms achieve small standard deviations for the Gaussian error model of $\sigma(\zeta),\sigma(\theta) < 0.016 \pi$, and small depolarizing error $p < 0.032 \, \%$, we can compare both schemes on a device. If the depolarizing error and the coherent error increase, we achieve a fidelity difference $\Delta \mathcal{F}^{de,co} \approx 0.3 \, \%$ (fig. \ref{fig:de_coherentczcp}; purple  region). In fig. \ref{fig:de_coherentczcp}$, \mathcal{F}_{cp}^{co,de}\approx 99.8 \, \%$ for the whole error range, whereas $\mathcal{F}_{cz}^{co,de}$ declines to $99.5 \, \%$ (top-right corner in fig. \ref{fig:de_coherentczcp}) for large error rates. 
    Therefore, the advantage of having a \textsc{CP} decomposition over CZ for large error rates is numerically proven. By increasing the incoherent error $p> 0.1 \, \%$, the dominance of the linear scaling law of the depolarizing error over the squared scaling of the coherent error would be identified. Thus the heatmap would change from a radial trend caused by the influence of $\sigma(\zeta),\sigma(\theta)$ and $p$ on the gate fidelities to a linear scaling in $p$ direction(y-axis).

   \section{Conclusion}
    We showed that the decomposition of the $ZZ$-gate into \textsc{CP} gate achieves greater gate fidelities compared to the \textsc{CZ}-gate decomposition in both incoherent and coherent error channels. By suppressing the depolarizing error below $0.03 \, \%$ and having a coherent error in the range of $1 \, \%$, both gate decompositions deliver comparable fidelities and could be used especially in the case of variational algorithms. The variational algorithm could deal with the coherent error due to the optimization process. For a coherent over-rotation $\theta > 0.04 \pi$, the gate fidelity in the \textsc{CZ} decomposition depends on the rotation angle $\gamma$. For $\zeta(\theta),\sigma(\theta) = 0.054 \pi$, the gate fidelity differs by $0.1 \, \%$,  which is significant concerning the quantum threshold theorem where a difference around $0.1 \, \%$ could make a difference of 1000 physical qubits per logical qubit \cite{threshold_2}. A fixed two-qubit gate like \textsc{CZ} in combination with single-qubit gates designed by virtual $Z$ gates could lead to a significant calibration reduction and less errors on pulse level. When the coherent over-rotation angle is below $\theta < 0.016 \pi$ ($0.8 \, \%$) and the incoherent error is suppressed below $0.03 \, \%$, we recommend \textsc{CZ} with virtual $Z$ gates.
    
    \section*{Acknowledgments}
   The authors acknowledge funding, support, and computational resources from German Aerospace Center (DLR) and the Forschungszentrum J{ü}lich. Furthermore, we acknowledge funding from QSolid funded by the Federal Ministry of Education and Research (BMBF). We also acknowledge funding from AQUAS and QUASIM, both funded by the Federal Ministry of Economic Affairs and Climate Action (BMWK). We acknowledge useful conversations with Alessandro Ciani and Tim Bode.
   
\input{main.bbl}

\end{document}

%% file: main.bbl
%

%% file: main.bbl
\begin{thebibliography}{32}%
\makeatletter
\providecommand \@ifxundefined [1]{%
 \@ifx{#1\undefined}
}%
\providecommand \@ifnum [1]{%
 \ifnum #1\expandafter \@firstoftwo
 \else \expandafter \@secondoftwo
 \fi
}%
\providecommand \@ifx [1]{%
 \ifx #1\expandafter \@firstoftwo
 \else \expandafter \@secondoftwo
 \fi
}%
\providecommand \natexlab [1]{#1}%
\providecommand \enquote  [1]{``#1''}%
\providecommand \bibnamefont  [1]{#1}%
\providecommand \bibfnamefont [1]{#1}%
\providecommand \citenamefont [1]{#1}%
\providecommand \href@noop [0]{\@secondoftwo}%
\providecommand \href [0]{\begingroup \@sanitize@url \@href}%
\providecommand \@href[1]{\@@startlink{#1}\@@href}%
\providecommand \@@href[1]{\endgroup#1\@@endlink}%
\providecommand \@sanitize@url [0]{\catcode `\\12\catcode `\$12\catcode
  `\&12\catcode `\#12\catcode `\^12\catcode `\_12\catcode `\%12\relax}%
\providecommand \@@startlink[1]{}%
\providecommand \@@endlink[0]{}%
\providecommand \url  [0]{\begingroup\@sanitize@url \@url }%
\providecommand \@url [1]{\endgroup\@href {#1}{\urlprefix }}%
\providecommand \urlprefix  [0]{URL }%
\providecommand \Eprint [0]{\href }%
\providecommand \doibase [0]{https://doi.org/}%
\providecommand \selectlanguage [0]{\@gobble}%
\providecommand \bibinfo  [0]{\@secondoftwo}%
\providecommand \bibfield  [0]{\@secondoftwo}%
\providecommand \translation [1]{[#1]}%
\providecommand \BibitemOpen [0]{}%
\providecommand \bibitemStop [0]{}%
\providecommand \bibitemNoStop [0]{.\EOS\space}%
\providecommand \EOS [0]{\spacefactor3000\relax}%
\providecommand \BibitemShut  [1]{\csname bibitem#1\endcsname}%
\let\auto@bib@innerbib\@empty
\bibitem [{\citenamefont {Shor}(1997)}]{Shor}%
  \BibitemOpen
  \bibfield  {author} {\bibinfo {author} {\bibfnamefont {P.~W.}\ \bibnamefont
  {Shor}},\ }\href@noop {} {\bibfield  {journal} {\bibinfo  {journal} {SIAM J.
  Comput.}\ }\textbf {\bibinfo {volume} {26}},\ \bibinfo {pages} {1484–1509}
  (\bibinfo {year} {1997})}\BibitemShut {NoStop}%
\bibitem [{\citenamefont {Grover}(1996)}]{Grover_first}%
  \BibitemOpen
  \bibfield  {author} {\bibinfo {author} {\bibfnamefont {L.~K.}\ \bibnamefont
  {Grover}},\ }in\ \href@noop {} {\emph {\bibinfo {booktitle} {Proceedings of
  the Twenty-Eighth Annual ACM Symposium on Theory of Computing}}}\ (\bibinfo
  {publisher} {Association for Computing Machinery},\ \bibinfo {address} {New
  York, NY, USA},\ \bibinfo {year} {1996})\ p.\ \bibinfo {pages}
  {212–219}\BibitemShut {NoStop}%
\bibitem [{\citenamefont {Arute}\ \emph {et~al.}(2019)\citenamefont {Arute},
  \citenamefont {Arya}, \citenamefont {Babbush}, \citenamefont {Bacon},
  \citenamefont {Bardin}, \citenamefont {Barends}, \citenamefont {Biswas},
  \citenamefont {Boixo}, \citenamefont {Brandao}, \citenamefont {Buell},
  \citenamefont {Burkett}, \citenamefont {Chen}, \citenamefont {Chen},
  \citenamefont {Chiaro}, \citenamefont {Collins}, \citenamefont {Courtney},
  \citenamefont {Dunsworth}, \citenamefont {Farhi}, \citenamefont {Foxen},
  \citenamefont {Fowler}, \citenamefont {Gidney}, \citenamefont {Giustina},
  \citenamefont {Graff}, \citenamefont {Guerin}, \citenamefont {Habegger},
  \citenamefont {Harrigan}, \citenamefont {Hartmann}, \citenamefont {Ho},
  \citenamefont {Hoffmann}, \citenamefont {Huang}, \citenamefont {Humble},
  \citenamefont {Isakov}, \citenamefont {Jeffrey}, \citenamefont {Jiang},
  \citenamefont {Kafri}, \citenamefont {Kechedzhi}, \citenamefont {Kelly},
  \citenamefont {Klimov}, \citenamefont {Knysh}, \citenamefont {Korotkov},
  \citenamefont {Kostritsa}, \citenamefont {Landhuis}, \citenamefont
  {Lindmark}, \citenamefont {Lucero}, \citenamefont {Lyakh}, \citenamefont
  {Mandr{\`a}}, \citenamefont {McClean}, \citenamefont {McEwen}, \citenamefont
  {Megrant}, \citenamefont {Mi}, \citenamefont {Michielsen}, \citenamefont
  {Mohseni}, \citenamefont {Mutus}, \citenamefont {Naaman}, \citenamefont
  {Neeley}, \citenamefont {Neill}, \citenamefont {Niu}, \citenamefont {Ostby},
  \citenamefont {Petukhov}, \citenamefont {Platt}, \citenamefont {Quintana},
  \citenamefont {Rieffel}, \citenamefont {Roushan}, \citenamefont {Rubin},
  \citenamefont {Sank}, \citenamefont {Satzinger}, \citenamefont {Smelyanskiy},
  \citenamefont {Sung}, \citenamefont {Trevithick}, \citenamefont
  {Vainsencher}, \citenamefont {Villalonga}, \citenamefont {White},
  \citenamefont {Yao}, \citenamefont {Yeh}, \citenamefont {Zalcman},
  \citenamefont {Neven},\ and\ \citenamefont {Martinis}}]{supremacy}%
  \BibitemOpen
  \bibfield  {author} {\bibinfo {author} {\bibfnamefont {F.}~\bibnamefont
  {Arute}}, \bibinfo {author} {\bibfnamefont {K.}~\bibnamefont {Arya}},
  \bibinfo {author} {\bibfnamefont {R.}~\bibnamefont {Babbush}}, \bibinfo
  {author} {\bibfnamefont {D.}~\bibnamefont {Bacon}}, \bibinfo {author}
  {\bibfnamefont {J.~C.}\ \bibnamefont {Bardin}}, \bibinfo {author}
  {\bibfnamefont {R.}~\bibnamefont {Barends}}, \bibinfo {author} {\bibfnamefont
  {R.}~\bibnamefont {Biswas}}, \bibinfo {author} {\bibfnamefont
  {S.}~\bibnamefont {Boixo}}, \bibinfo {author} {\bibfnamefont {F.~G. S.~L.}\
  \bibnamefont {Brandao}}, \bibinfo {author} {\bibfnamefont {D.~A.}\
  \bibnamefont {Buell}}, \bibinfo {author} {\bibfnamefont {B.}~\bibnamefont
  {Burkett}}, \bibinfo {author} {\bibfnamefont {Y.}~\bibnamefont {Chen}},
  \bibinfo {author} {\bibfnamefont {Z.}~\bibnamefont {Chen}}, \bibinfo {author}
  {\bibfnamefont {B.}~\bibnamefont {Chiaro}}, \bibinfo {author} {\bibfnamefont
  {R.}~\bibnamefont {Collins}}, \bibinfo {author} {\bibfnamefont
  {W.}~\bibnamefont {Courtney}}, \bibinfo {author} {\bibfnamefont
  {A.}~\bibnamefont {Dunsworth}}, \bibinfo {author} {\bibfnamefont
  {E.}~\bibnamefont {Farhi}}, \bibinfo {author} {\bibfnamefont
  {B.}~\bibnamefont {Foxen}}, \bibinfo {author} {\bibfnamefont
  {A.}~\bibnamefont {Fowler}}, \bibinfo {author} {\bibfnamefont
  {C.}~\bibnamefont {Gidney}}, \bibinfo {author} {\bibfnamefont
  {M.}~\bibnamefont {Giustina}}, \bibinfo {author} {\bibfnamefont
  {R.}~\bibnamefont {Graff}}, \bibinfo {author} {\bibfnamefont
  {K.}~\bibnamefont {Guerin}}, \bibinfo {author} {\bibfnamefont
  {S.}~\bibnamefont {Habegger}}, \bibinfo {author} {\bibfnamefont {M.~P.}\
  \bibnamefont {Harrigan}}, \bibinfo {author} {\bibfnamefont {M.~J.}\
  \bibnamefont {Hartmann}}, \bibinfo {author} {\bibfnamefont {A.}~\bibnamefont
  {Ho}}, \bibinfo {author} {\bibfnamefont {M.}~\bibnamefont {Hoffmann}},
  \bibinfo {author} {\bibfnamefont {T.}~\bibnamefont {Huang}}, \bibinfo
  {author} {\bibfnamefont {T.~S.}\ \bibnamefont {Humble}}, \bibinfo {author}
  {\bibfnamefont {S.~V.}\ \bibnamefont {Isakov}}, \bibinfo {author}
  {\bibfnamefont {E.}~\bibnamefont {Jeffrey}}, \bibinfo {author} {\bibfnamefont
  {Z.}~\bibnamefont {Jiang}}, \bibinfo {author} {\bibfnamefont
  {D.}~\bibnamefont {Kafri}}, \bibinfo {author} {\bibfnamefont
  {K.}~\bibnamefont {Kechedzhi}}, \bibinfo {author} {\bibfnamefont
  {J.}~\bibnamefont {Kelly}}, \bibinfo {author} {\bibfnamefont {P.~V.}\
  \bibnamefont {Klimov}}, \bibinfo {author} {\bibfnamefont {S.}~\bibnamefont
  {Knysh}}, \bibinfo {author} {\bibfnamefont {A.}~\bibnamefont {Korotkov}},
  \bibinfo {author} {\bibfnamefont {F.}~\bibnamefont {Kostritsa}}, \bibinfo
  {author} {\bibfnamefont {D.}~\bibnamefont {Landhuis}}, \bibinfo {author}
  {\bibfnamefont {M.}~\bibnamefont {Lindmark}}, \bibinfo {author}
  {\bibfnamefont {E.}~\bibnamefont {Lucero}}, \bibinfo {author} {\bibfnamefont
  {D.}~\bibnamefont {Lyakh}}, \bibinfo {author} {\bibfnamefont
  {S.}~\bibnamefont {Mandr{\`a}}}, \bibinfo {author} {\bibfnamefont {J.~R.}\
  \bibnamefont {McClean}}, \bibinfo {author} {\bibfnamefont {M.}~\bibnamefont
  {McEwen}}, \bibinfo {author} {\bibfnamefont {A.}~\bibnamefont {Megrant}},
  \bibinfo {author} {\bibfnamefont {X.}~\bibnamefont {Mi}}, \bibinfo {author}
  {\bibfnamefont {K.}~\bibnamefont {Michielsen}}, \bibinfo {author}
  {\bibfnamefont {M.}~\bibnamefont {Mohseni}}, \bibinfo {author} {\bibfnamefont
  {J.}~\bibnamefont {Mutus}}, \bibinfo {author} {\bibfnamefont
  {O.}~\bibnamefont {Naaman}}, \bibinfo {author} {\bibfnamefont
  {M.}~\bibnamefont {Neeley}}, \bibinfo {author} {\bibfnamefont
  {C.}~\bibnamefont {Neill}}, \bibinfo {author} {\bibfnamefont {M.~Y.}\
  \bibnamefont {Niu}}, \bibinfo {author} {\bibfnamefont {E.}~\bibnamefont
  {Ostby}}, \bibinfo {author} {\bibfnamefont {A.}~\bibnamefont {Petukhov}},
  \bibinfo {author} {\bibfnamefont {J.~C.}\ \bibnamefont {Platt}}, \bibinfo
  {author} {\bibfnamefont {C.}~\bibnamefont {Quintana}}, \bibinfo {author}
  {\bibfnamefont {E.~G.}\ \bibnamefont {Rieffel}}, \bibinfo {author}
  {\bibfnamefont {P.}~\bibnamefont {Roushan}}, \bibinfo {author} {\bibfnamefont
  {N.~C.}\ \bibnamefont {Rubin}}, \bibinfo {author} {\bibfnamefont
  {D.}~\bibnamefont {Sank}}, \bibinfo {author} {\bibfnamefont {K.~J.}\
  \bibnamefont {Satzinger}}, \bibinfo {author} {\bibfnamefont {V.}~\bibnamefont
  {Smelyanskiy}}, \bibinfo {author} {\bibfnamefont {K.~J.}\ \bibnamefont
  {Sung}}, \bibinfo {author} {\bibfnamefont {M.~D.}\ \bibnamefont
  {Trevithick}}, \bibinfo {author} {\bibfnamefont {A.}~\bibnamefont
  {Vainsencher}}, \bibinfo {author} {\bibfnamefont {B.}~\bibnamefont
  {Villalonga}}, \bibinfo {author} {\bibfnamefont {T.}~\bibnamefont {White}},
  \bibinfo {author} {\bibfnamefont {Z.~J.}\ \bibnamefont {Yao}}, \bibinfo
  {author} {\bibfnamefont {P.}~\bibnamefont {Yeh}}, \bibinfo {author}
  {\bibfnamefont {A.}~\bibnamefont {Zalcman}}, \bibinfo {author} {\bibfnamefont
  {H.}~\bibnamefont {Neven}},\ and\ \bibinfo {author} {\bibfnamefont {J.~M.}\
  \bibnamefont {Martinis}},\ }\href@noop {} {\bibfield  {journal} {\bibinfo
  {journal} {Nature}\ }\textbf {\bibinfo {volume} {574}},\ \bibinfo {pages}
  {505} (\bibinfo {year} {2019})}\BibitemShut {NoStop}%
\bibitem [{\citenamefont {Barends}\ \emph {et~al.}(2014)\citenamefont
  {Barends}, \citenamefont {Kelly}, \citenamefont {Megrant}, \citenamefont
  {Veitia}, \citenamefont {Sank}, \citenamefont {Jeffrey}, \citenamefont
  {White}, \citenamefont {Mutus}, \citenamefont {Fowler}, \citenamefont
  {Campbell}, \citenamefont {Chen}, \citenamefont {Chen}, \citenamefont
  {Chiaro}, \citenamefont {Dunsworth}, \citenamefont {Neill}, \citenamefont
  {O'Malley}, \citenamefont {Roushan}, \citenamefont {Vainsencher},
  \citenamefont {Wenner}, \citenamefont {Korotkov}, \citenamefont {Cleland},\
  and\ \citenamefont {Martinis}}]{Barends_Martinis}%
  \BibitemOpen
  \bibfield  {author} {\bibinfo {author} {\bibfnamefont {R.}~\bibnamefont
  {Barends}}, \bibinfo {author} {\bibfnamefont {J.}~\bibnamefont {Kelly}},
  \bibinfo {author} {\bibfnamefont {A.}~\bibnamefont {Megrant}}, \bibinfo
  {author} {\bibfnamefont {A.}~\bibnamefont {Veitia}}, \bibinfo {author}
  {\bibfnamefont {D.}~\bibnamefont {Sank}}, \bibinfo {author} {\bibfnamefont
  {E.}~\bibnamefont {Jeffrey}}, \bibinfo {author} {\bibfnamefont {T.~C.}\
  \bibnamefont {White}}, \bibinfo {author} {\bibfnamefont {J.}~\bibnamefont
  {Mutus}}, \bibinfo {author} {\bibfnamefont {A.~G.}\ \bibnamefont {Fowler}},
  \bibinfo {author} {\bibfnamefont {B.}~\bibnamefont {Campbell}}, \bibinfo
  {author} {\bibfnamefont {Y.}~\bibnamefont {Chen}}, \bibinfo {author}
  {\bibfnamefont {Z.}~\bibnamefont {Chen}}, \bibinfo {author} {\bibfnamefont
  {B.}~\bibnamefont {Chiaro}}, \bibinfo {author} {\bibfnamefont
  {A.}~\bibnamefont {Dunsworth}}, \bibinfo {author} {\bibfnamefont
  {C.}~\bibnamefont {Neill}}, \bibinfo {author} {\bibfnamefont
  {P.}~\bibnamefont {O'Malley}}, \bibinfo {author} {\bibfnamefont
  {P.}~\bibnamefont {Roushan}}, \bibinfo {author} {\bibfnamefont
  {A.}~\bibnamefont {Vainsencher}}, \bibinfo {author} {\bibfnamefont
  {J.}~\bibnamefont {Wenner}}, \bibinfo {author} {\bibfnamefont {A.~N.}\
  \bibnamefont {Korotkov}}, \bibinfo {author} {\bibfnamefont {A.~N.}\
  \bibnamefont {Cleland}},\ and\ \bibinfo {author} {\bibfnamefont {J.~M.}\
  \bibnamefont {Martinis}},\ }\href@noop {} {\bibfield  {journal} {\bibinfo
  {journal} {Nature}\ }\textbf {\bibinfo {volume} {508}},\ \bibinfo {pages}
  {500} (\bibinfo {year} {2014})}\BibitemShut {NoStop}%
\bibitem [{\citenamefont {Farhi}\ \emph {et~al.}(2014)\citenamefont {Farhi},
  \citenamefont {Goldstone},\ and\ \citenamefont {Gutmann}}]{farhi2014quantum}%
  \BibitemOpen
  \bibfield  {author} {\bibinfo {author} {\bibfnamefont {E.}~\bibnamefont
  {Farhi}}, \bibinfo {author} {\bibfnamefont {J.}~\bibnamefont {Goldstone}},\
  and\ \bibinfo {author} {\bibfnamefont {S.}~\bibnamefont {Gutmann}},\
  }\href@noop {} {\bibinfo {title} {A quantum approximate optimization
  algorithm}} (\bibinfo {year} {2014}),\ \Eprint
  {https://arxiv.org/abs/1411.4028} {arXiv:1411.4028 [quant-ph]} \BibitemShut
  {NoStop}%
\bibitem [{\citenamefont {Peruzzo}\ \emph {et~al.}(2014)\citenamefont
  {Peruzzo}, \citenamefont {McClean}, \citenamefont {Shadbolt}, \citenamefont
  {Yung}, \citenamefont {Zhou}, \citenamefont {Love}, \citenamefont
  {Aspuru-Guzik},\ and\ \citenamefont {O'Brien}}]{Peruzzo2014}%
  \BibitemOpen
  \bibfield  {author} {\bibinfo {author} {\bibfnamefont {A.}~\bibnamefont
  {Peruzzo}}, \bibinfo {author} {\bibfnamefont {J.}~\bibnamefont {McClean}},
  \bibinfo {author} {\bibfnamefont {P.}~\bibnamefont {Shadbolt}}, \bibinfo
  {author} {\bibfnamefont {M.-H.}\ \bibnamefont {Yung}}, \bibinfo {author}
  {\bibfnamefont {X.-Q.}\ \bibnamefont {Zhou}}, \bibinfo {author}
  {\bibfnamefont {P.~J.}\ \bibnamefont {Love}}, \bibinfo {author}
  {\bibfnamefont {A.}~\bibnamefont {Aspuru-Guzik}},\ and\ \bibinfo {author}
  {\bibfnamefont {J.~L.}\ \bibnamefont {O'Brien}},\ }\href@noop {} {\bibfield
  {journal} {\bibinfo  {journal} {Nature Communications}\ }\textbf {\bibinfo
  {volume} {5}},\ \bibinfo {pages} {4213} (\bibinfo {year} {2014})}\BibitemShut
  {NoStop}%
\bibitem [{\citenamefont {Suzuki}\ \emph {et~al.}(2022)\citenamefont {Suzuki},
  \citenamefont {Endo}, \citenamefont {Fujii},\ and\ \citenamefont
  {Tokunaga}}]{FTQC_limits}%
  \BibitemOpen
  \bibfield  {author} {\bibinfo {author} {\bibfnamefont {Y.}~\bibnamefont
  {Suzuki}}, \bibinfo {author} {\bibfnamefont {S.}~\bibnamefont {Endo}},
  \bibinfo {author} {\bibfnamefont {K.}~\bibnamefont {Fujii}},\ and\ \bibinfo
  {author} {\bibfnamefont {Y.}~\bibnamefont {Tokunaga}},\ }\href@noop {}
  {\bibfield  {journal} {\bibinfo  {journal} {PRX Quantum}\ }\textbf {\bibinfo
  {volume} {3}},\ \bibinfo {pages} {010345} (\bibinfo {year}
  {2022})}\BibitemShut {NoStop}%
\bibitem [{\citenamefont {Farhi}\ and\ \citenamefont
  {Harrow}(2019)}]{supremacy_QAOA}%
  \BibitemOpen
  \bibfield  {author} {\bibinfo {author} {\bibfnamefont {E.}~\bibnamefont
  {Farhi}}\ and\ \bibinfo {author} {\bibfnamefont {A.~W.}\ \bibnamefont
  {Harrow}},\ }\href@noop {} {\bibinfo {title} {Quantum supremacy through the
  quantum approximate optimization algorithm}} (\bibinfo {year} {2019}),\
  \Eprint {https://arxiv.org/abs/1602.07674} {arXiv:1602.07674 [quant-ph]}
  \BibitemShut {NoStop}%
\bibitem [{\citenamefont {Zhou}\ \emph {et~al.}(2020)\citenamefont {Zhou},
  \citenamefont {Wang}, \citenamefont {Choi}, \citenamefont {Pichler},\ and\
  \citenamefont {Lukin}}]{qaoa_analytic_sim}%
  \BibitemOpen
  \bibfield  {author} {\bibinfo {author} {\bibfnamefont {L.}~\bibnamefont
  {Zhou}}, \bibinfo {author} {\bibfnamefont {S.-T.}\ \bibnamefont {Wang}},
  \bibinfo {author} {\bibfnamefont {S.}~\bibnamefont {Choi}}, \bibinfo {author}
  {\bibfnamefont {H.}~\bibnamefont {Pichler}},\ and\ \bibinfo {author}
  {\bibfnamefont {M.~D.}\ \bibnamefont {Lukin}},\ }\href@noop {} {\bibfield
  {journal} {\bibinfo  {journal} {Phys. Rev. X}\ }\textbf {\bibinfo {volume}
  {10}},\ \bibinfo {pages} {021067} (\bibinfo {year} {2020})}\BibitemShut
  {NoStop}%
\bibitem [{\citenamefont {Streif}\ and\ \citenamefont
  {Leib}(2020)}]{qaoa_analytic}%
  \BibitemOpen
  \bibfield  {author} {\bibinfo {author} {\bibfnamefont {M.}~\bibnamefont
  {Streif}}\ and\ \bibinfo {author} {\bibfnamefont {M.}~\bibnamefont {Leib}},\
  }\href@noop {} {\bibfield  {journal} {\bibinfo  {journal} {Quantum Science
  and Technology}\ }\textbf {\bibinfo {volume} {5}},\ \bibinfo {pages} {034008}
  (\bibinfo {year} {2020})}\BibitemShut {NoStop}%
\bibitem [{\citenamefont {Akshay}\ \emph {et~al.}(2020)\citenamefont {Akshay},
  \citenamefont {Philathong}, \citenamefont {Morales},\ and\ \citenamefont
  {Biamonte}}]{qaoa_analytic_II}%
  \BibitemOpen
  \bibfield  {author} {\bibinfo {author} {\bibfnamefont {V.}~\bibnamefont
  {Akshay}}, \bibinfo {author} {\bibfnamefont {H.}~\bibnamefont {Philathong}},
  \bibinfo {author} {\bibfnamefont {M.~E.~S.}\ \bibnamefont {Morales}},\ and\
  \bibinfo {author} {\bibfnamefont {J.~D.}\ \bibnamefont {Biamonte}},\
  }\href@noop {} {\bibfield  {journal} {\bibinfo  {journal} {Phys. Rev. Lett.}\
  }\textbf {\bibinfo {volume} {124}},\ \bibinfo {pages} {090504} (\bibinfo
  {year} {2020})}\BibitemShut {NoStop}%
\bibitem [{\citenamefont {Amin}()}]{transverse_field}%
  \BibitemOpen
  \bibfield  {author} {\bibinfo {author} {\bibfnamefont {M.~H.~S.}\
  \bibnamefont {Amin}},\ }\href@noop {} {\bibinfo  {journal} {Phys. Rev.
  Lett.}\ }\BibitemShut {NoStop}%
\bibitem [{\citenamefont {Streif}\ \emph {et~al.}(2020)\citenamefont {Streif},
  \citenamefont {Leib}, \citenamefont {Wudarski}, \citenamefont {Rieffel},\
  and\ \citenamefont {Wang}}]{streif2020quantum}%
  \BibitemOpen
\bibfield  {journal} {  }\bibfield  {author} {\bibinfo {author} {\bibfnamefont
  {M.}~\bibnamefont {Streif}}, \bibinfo {author} {\bibfnamefont
  {M.}~\bibnamefont {Leib}}, \bibinfo {author} {\bibfnamefont {F.}~\bibnamefont
  {Wudarski}}, \bibinfo {author} {\bibfnamefont {E.}~\bibnamefont {Rieffel}},\
  and\ \bibinfo {author} {\bibfnamefont {Z.}~\bibnamefont {Wang}},\ }\href@noop
  {} {\bibinfo {title} {Quantum algorithms with local particle number
  conservation: noise effects and error correction}} (\bibinfo {year} {2020}),\
  \Eprint {https://arxiv.org/abs/2011.06873} {arXiv:2011.06873 [quant-ph]}
  \BibitemShut {NoStop}%
\bibitem [{\citenamefont {Marshall}\ \emph {et~al.}(2020)\citenamefont
  {Marshall}, \citenamefont {Wudarski}, \citenamefont {Hadfield},\ and\
  \citenamefont {Hogg}}]{Marshall_2020}%
  \BibitemOpen
  \bibfield  {author} {\bibinfo {author} {\bibfnamefont {J.}~\bibnamefont
  {Marshall}}, \bibinfo {author} {\bibfnamefont {F.}~\bibnamefont {Wudarski}},
  \bibinfo {author} {\bibfnamefont {S.}~\bibnamefont {Hadfield}},\ and\
  \bibinfo {author} {\bibfnamefont {T.}~\bibnamefont {Hogg}},\ }\href@noop {}
  {\bibfield  {journal} {\bibinfo  {journal} {{IOP} {SciNotes}}\ }\textbf
  {\bibinfo {volume} {1}},\ \bibinfo {pages} {025208} (\bibinfo {year}
  {2020})}\BibitemShut {NoStop}%
\bibitem [{\citenamefont {Yang}\ \emph {et~al.}(2019)\citenamefont {Yang},
  \citenamefont {Coppersmith},\ and\ \citenamefont {Friesen}}]{charge_noise}%
  \BibitemOpen
  \bibfield  {author} {\bibinfo {author} {\bibfnamefont {Y.-C.}\ \bibnamefont
  {Yang}}, \bibinfo {author} {\bibfnamefont {S.~N.}\ \bibnamefont
  {Coppersmith}},\ and\ \bibinfo {author} {\bibfnamefont {M.}~\bibnamefont
  {Friesen}},\ }\href@noop {} {\bibfield  {journal} {\bibinfo  {journal} {npj
  Quantum Information}\ }\textbf {\bibinfo {volume} {5}},\ \bibinfo {pages}
  {12} (\bibinfo {year} {2019})}\BibitemShut {NoStop}%
\bibitem [{\citenamefont {Tosi}\ \emph {et~al.}(2017)\citenamefont {Tosi},
  \citenamefont {Mohiyaddin}, \citenamefont {Schmitt}, \citenamefont {Tenberg},
  \citenamefont {Rahman}, \citenamefont {Klimeck},\ and\ \citenamefont
  {Morello}}]{Tosi2017}%
  \BibitemOpen
  \bibfield  {author} {\bibinfo {author} {\bibfnamefont {G.}~\bibnamefont
  {Tosi}}, \bibinfo {author} {\bibfnamefont {F.~A.}\ \bibnamefont
  {Mohiyaddin}}, \bibinfo {author} {\bibfnamefont {V.}~\bibnamefont {Schmitt}},
  \bibinfo {author} {\bibfnamefont {S.}~\bibnamefont {Tenberg}}, \bibinfo
  {author} {\bibfnamefont {R.}~\bibnamefont {Rahman}}, \bibinfo {author}
  {\bibfnamefont {G.}~\bibnamefont {Klimeck}},\ and\ \bibinfo {author}
  {\bibfnamefont {A.}~\bibnamefont {Morello}},\ }\href@noop {} {\bibfield
  {journal} {\bibinfo  {journal} {Nature Communications}\ }\textbf {\bibinfo
  {volume} {8}},\ \bibinfo {pages} {450} (\bibinfo {year} {2017})}\BibitemShut
  {NoStop}%
\bibitem [{\citenamefont {Li}\ \emph {et~al.}(2020)\citenamefont {Li},
  \citenamefont {Cai}, \citenamefont {Yan}, \citenamefont {Wang}, \citenamefont
  {Pan}, \citenamefont {Ma}, \citenamefont {Cai}, \citenamefont {Han},
  \citenamefont {Hua}, \citenamefont {Han}, \citenamefont {Wu}, \citenamefont
  {Zhang}, \citenamefont {Wang}, \citenamefont {Song}, \citenamefont {Duan},\
  and\ \citenamefont {Sun}}]{tubale_coupler}%
  \BibitemOpen
  \bibfield  {author} {\bibinfo {author} {\bibfnamefont {X.}~\bibnamefont
  {Li}}, \bibinfo {author} {\bibfnamefont {T.}~\bibnamefont {Cai}}, \bibinfo
  {author} {\bibfnamefont {H.}~\bibnamefont {Yan}}, \bibinfo {author}
  {\bibfnamefont {Z.}~\bibnamefont {Wang}}, \bibinfo {author} {\bibfnamefont
  {X.}~\bibnamefont {Pan}}, \bibinfo {author} {\bibfnamefont {Y.}~\bibnamefont
  {Ma}}, \bibinfo {author} {\bibfnamefont {W.}~\bibnamefont {Cai}}, \bibinfo
  {author} {\bibfnamefont {J.}~\bibnamefont {Han}}, \bibinfo {author}
  {\bibfnamefont {Z.}~\bibnamefont {Hua}}, \bibinfo {author} {\bibfnamefont
  {X.}~\bibnamefont {Han}}, \bibinfo {author} {\bibfnamefont {Y.}~\bibnamefont
  {Wu}}, \bibinfo {author} {\bibfnamefont {H.}~\bibnamefont {Zhang}}, \bibinfo
  {author} {\bibfnamefont {H.}~\bibnamefont {Wang}}, \bibinfo {author}
  {\bibfnamefont {Y.}~\bibnamefont {Song}}, \bibinfo {author} {\bibfnamefont
  {L.}~\bibnamefont {Duan}},\ and\ \bibinfo {author} {\bibfnamefont
  {L.}~\bibnamefont {Sun}},\ }\href@noop {} {\bibfield  {journal} {\bibinfo
  {journal} {Phys. Rev. Applied}\ }\textbf {\bibinfo {volume} {14}},\ \bibinfo
  {pages} {024070} (\bibinfo {year} {2020})}\BibitemShut {NoStop}%
\bibitem [{\citenamefont {Linke}\ \emph {et~al.}(2017)\citenamefont {Linke},
  \citenamefont {Maslov}, \citenamefont {Roetteler}, \citenamefont {Debnath},
  \citenamefont {Figgatt}, \citenamefont {Landsman}, \citenamefont {Wright},\
  and\ \citenamefont {Monroe}}]{speed_ratio}%
  \BibitemOpen
  \bibfield  {author} {\bibinfo {author} {\bibfnamefont {N.~M.}\ \bibnamefont
  {Linke}}, \bibinfo {author} {\bibfnamefont {D.}~\bibnamefont {Maslov}},
  \bibinfo {author} {\bibfnamefont {M.}~\bibnamefont {Roetteler}}, \bibinfo
  {author} {\bibfnamefont {S.}~\bibnamefont {Debnath}}, \bibinfo {author}
  {\bibfnamefont {C.}~\bibnamefont {Figgatt}}, \bibinfo {author} {\bibfnamefont
  {K.~A.}\ \bibnamefont {Landsman}}, \bibinfo {author} {\bibfnamefont
  {K.}~\bibnamefont {Wright}},\ and\ \bibinfo {author} {\bibfnamefont
  {C.}~\bibnamefont {Monroe}},\ }\href@noop {} {\bibfield  {journal} {\bibinfo
  {journal} {Proceedings of the National Academy of Sciences}\ }\textbf
  {\bibinfo {volume} {114}},\ \bibinfo {pages} {3305} (\bibinfo {year}
  {2017})}\BibitemShut {NoStop}%
\bibitem [{\citenamefont {Foxen}\ \emph {et~al.}(2020)\citenamefont {Foxen},
  \citenamefont {Neill}, \citenamefont {Dunsworth}, \citenamefont {Roushan},
  \citenamefont {Chiaro}, \citenamefont {Megrant}, \citenamefont {Kelly},
  \citenamefont {Chen}, \citenamefont {Satzinger}, \citenamefont {Barends},
  \citenamefont {Arute}, \citenamefont {Arya}, \citenamefont {Babbush},
  \citenamefont {Bacon}, \citenamefont {Bardin}, \citenamefont {Boixo},
  \citenamefont {Buell}, \citenamefont {Burkett}, \citenamefont {Chen},
  \citenamefont {Collins}, \citenamefont {Farhi}, \citenamefont {Fowler},
  \citenamefont {Gidney}, \citenamefont {Giustina}, \citenamefont {Graff},
  \citenamefont {Harrigan}, \citenamefont {Huang}, \citenamefont {Isakov},
  \citenamefont {Jeffrey}, \citenamefont {Jiang}, \citenamefont {Kafri},
  \citenamefont {Kechedzhi}, \citenamefont {Klimov}, \citenamefont {Korotkov},
  \citenamefont {Kostritsa}, \citenamefont {Landhuis}, \citenamefont {Lucero},
  \citenamefont {McClean}, \citenamefont {McEwen}, \citenamefont {Mi},
  \citenamefont {Mohseni}, \citenamefont {Mutus}, \citenamefont {Naaman},
  \citenamefont {Neeley}, \citenamefont {Niu}, \citenamefont {Petukhov},
  \citenamefont {Quintana}, \citenamefont {Rubin}, \citenamefont {Sank},
  \citenamefont {Smelyanskiy}, \citenamefont {Vainsencher}, \citenamefont
  {White}, \citenamefont {Yao}, \citenamefont {Yeh}, \citenamefont {Zalcman},
  \citenamefont {Neven},\ and\ \citenamefont {Martinis}}]{google_fsim}%
  \BibitemOpen
  \bibfield  {author} {\bibinfo {author} {\bibfnamefont {B.}~\bibnamefont
  {Foxen}}, \bibinfo {author} {\bibfnamefont {C.}~\bibnamefont {Neill}},
  \bibinfo {author} {\bibfnamefont {A.}~\bibnamefont {Dunsworth}}, \bibinfo
  {author} {\bibfnamefont {P.}~\bibnamefont {Roushan}}, \bibinfo {author}
  {\bibfnamefont {B.}~\bibnamefont {Chiaro}}, \bibinfo {author} {\bibfnamefont
  {A.}~\bibnamefont {Megrant}}, \bibinfo {author} {\bibfnamefont
  {J.}~\bibnamefont {Kelly}}, \bibinfo {author} {\bibfnamefont
  {Z.}~\bibnamefont {Chen}}, \bibinfo {author} {\bibfnamefont {K.}~\bibnamefont
  {Satzinger}}, \bibinfo {author} {\bibfnamefont {R.}~\bibnamefont {Barends}},
  \bibinfo {author} {\bibfnamefont {F.}~\bibnamefont {Arute}}, \bibinfo
  {author} {\bibfnamefont {K.}~\bibnamefont {Arya}}, \bibinfo {author}
  {\bibfnamefont {R.}~\bibnamefont {Babbush}}, \bibinfo {author} {\bibfnamefont
  {D.}~\bibnamefont {Bacon}}, \bibinfo {author} {\bibfnamefont {J.~C.}\
  \bibnamefont {Bardin}}, \bibinfo {author} {\bibfnamefont {S.}~\bibnamefont
  {Boixo}}, \bibinfo {author} {\bibfnamefont {D.}~\bibnamefont {Buell}},
  \bibinfo {author} {\bibfnamefont {B.}~\bibnamefont {Burkett}}, \bibinfo
  {author} {\bibfnamefont {Y.}~\bibnamefont {Chen}}, \bibinfo {author}
  {\bibfnamefont {R.}~\bibnamefont {Collins}}, \bibinfo {author} {\bibfnamefont
  {E.}~\bibnamefont {Farhi}}, \bibinfo {author} {\bibfnamefont
  {A.}~\bibnamefont {Fowler}}, \bibinfo {author} {\bibfnamefont
  {C.}~\bibnamefont {Gidney}}, \bibinfo {author} {\bibfnamefont
  {M.}~\bibnamefont {Giustina}}, \bibinfo {author} {\bibfnamefont
  {R.}~\bibnamefont {Graff}}, \bibinfo {author} {\bibfnamefont
  {M.}~\bibnamefont {Harrigan}}, \bibinfo {author} {\bibfnamefont
  {T.}~\bibnamefont {Huang}}, \bibinfo {author} {\bibfnamefont {S.~V.}\
  \bibnamefont {Isakov}}, \bibinfo {author} {\bibfnamefont {E.}~\bibnamefont
  {Jeffrey}}, \bibinfo {author} {\bibfnamefont {Z.}~\bibnamefont {Jiang}},
  \bibinfo {author} {\bibfnamefont {D.}~\bibnamefont {Kafri}}, \bibinfo
  {author} {\bibfnamefont {K.}~\bibnamefont {Kechedzhi}}, \bibinfo {author}
  {\bibfnamefont {P.}~\bibnamefont {Klimov}}, \bibinfo {author} {\bibfnamefont
  {A.}~\bibnamefont {Korotkov}}, \bibinfo {author} {\bibfnamefont
  {F.}~\bibnamefont {Kostritsa}}, \bibinfo {author} {\bibfnamefont
  {D.}~\bibnamefont {Landhuis}}, \bibinfo {author} {\bibfnamefont
  {E.}~\bibnamefont {Lucero}}, \bibinfo {author} {\bibfnamefont
  {J.}~\bibnamefont {McClean}}, \bibinfo {author} {\bibfnamefont
  {M.}~\bibnamefont {McEwen}}, \bibinfo {author} {\bibfnamefont
  {X.}~\bibnamefont {Mi}}, \bibinfo {author} {\bibfnamefont {M.}~\bibnamefont
  {Mohseni}}, \bibinfo {author} {\bibfnamefont {J.~Y.}\ \bibnamefont {Mutus}},
  \bibinfo {author} {\bibfnamefont {O.}~\bibnamefont {Naaman}}, \bibinfo
  {author} {\bibfnamefont {M.}~\bibnamefont {Neeley}}, \bibinfo {author}
  {\bibfnamefont {M.}~\bibnamefont {Niu}}, \bibinfo {author} {\bibfnamefont
  {A.}~\bibnamefont {Petukhov}}, \bibinfo {author} {\bibfnamefont
  {C.}~\bibnamefont {Quintana}}, \bibinfo {author} {\bibfnamefont
  {N.}~\bibnamefont {Rubin}}, \bibinfo {author} {\bibfnamefont
  {D.}~\bibnamefont {Sank}}, \bibinfo {author} {\bibfnamefont {V.}~\bibnamefont
  {Smelyanskiy}}, \bibinfo {author} {\bibfnamefont {A.}~\bibnamefont
  {Vainsencher}}, \bibinfo {author} {\bibfnamefont {T.~C.}\ \bibnamefont
  {White}}, \bibinfo {author} {\bibfnamefont {Z.}~\bibnamefont {Yao}}, \bibinfo
  {author} {\bibfnamefont {P.}~\bibnamefont {Yeh}}, \bibinfo {author}
  {\bibfnamefont {A.}~\bibnamefont {Zalcman}}, \bibinfo {author} {\bibfnamefont
  {H.}~\bibnamefont {Neven}},\ and\ \bibinfo {author} {\bibfnamefont {J.~M.}\
  \bibnamefont {Martinis}} (\bibinfo {collaboration} {Google AI Quantum}),\
  }\href@noop {} {\bibfield  {journal} {\bibinfo  {journal} {Phys. Rev. Lett.}\
  }\textbf {\bibinfo {volume} {125}},\ \bibinfo {pages} {120504} (\bibinfo
  {year} {2020})}\BibitemShut {NoStop}%
\bibitem [{\citenamefont {Abrams}\ \emph {et~al.}(2020)\citenamefont {Abrams},
  \citenamefont {Didier}, \citenamefont {Johnson}, \citenamefont {Silva},\ and\
  \citenamefont {Ryan}}]{rigetti}%
  \BibitemOpen
  \bibfield  {author} {\bibinfo {author} {\bibfnamefont {D.~M.}\ \bibnamefont
  {Abrams}}, \bibinfo {author} {\bibfnamefont {N.}~\bibnamefont {Didier}},
  \bibinfo {author} {\bibfnamefont {B.~R.}\ \bibnamefont {Johnson}}, \bibinfo
  {author} {\bibfnamefont {M.~P.~d.}\ \bibnamefont {Silva}},\ and\ \bibinfo
  {author} {\bibfnamefont {C.~A.}\ \bibnamefont {Ryan}},\ }\href@noop {}
  {\bibfield  {journal} {\bibinfo  {journal} {Nature Electronics}\ }\textbf
  {\bibinfo {volume} {3}},\ \bibinfo {pages} {744} (\bibinfo {year}
  {2020})}\BibitemShut {NoStop}%
\bibitem [{\citenamefont {Lacroix}\ \emph {et~al.}(2020)\citenamefont
  {Lacroix}, \citenamefont {Hellings}, \citenamefont {Andersen}, \citenamefont
  {Di~Paolo}, \citenamefont {Remm}, \citenamefont {Lazar}, \citenamefont
  {Krinner}, \citenamefont {Norris}, \citenamefont {Gabureac}, \citenamefont
  {Heinsoo}, \citenamefont {Blais}, \citenamefont {Eichler},\ and\
  \citenamefont {Wallraff}}]{Wallraff_cp_cz}%
  \BibitemOpen
  \bibfield  {author} {\bibinfo {author} {\bibfnamefont {N.}~\bibnamefont
  {Lacroix}}, \bibinfo {author} {\bibfnamefont {C.}~\bibnamefont {Hellings}},
  \bibinfo {author} {\bibfnamefont {C.~K.}\ \bibnamefont {Andersen}}, \bibinfo
  {author} {\bibfnamefont {A.}~\bibnamefont {Di~Paolo}}, \bibinfo {author}
  {\bibfnamefont {A.}~\bibnamefont {Remm}}, \bibinfo {author} {\bibfnamefont
  {S.}~\bibnamefont {Lazar}}, \bibinfo {author} {\bibfnamefont
  {S.}~\bibnamefont {Krinner}}, \bibinfo {author} {\bibfnamefont {G.~J.}\
  \bibnamefont {Norris}}, \bibinfo {author} {\bibfnamefont {M.}~\bibnamefont
  {Gabureac}}, \bibinfo {author} {\bibfnamefont {J.}~\bibnamefont {Heinsoo}},
  \bibinfo {author} {\bibfnamefont {A.}~\bibnamefont {Blais}}, \bibinfo
  {author} {\bibfnamefont {C.}~\bibnamefont {Eichler}},\ and\ \bibinfo {author}
  {\bibfnamefont {A.}~\bibnamefont {Wallraff}},\ }\href@noop {} {\bibfield
  {journal} {\bibinfo  {journal} {PRX Quantum}\ }\textbf {\bibinfo {volume}
  {1}},\ \bibinfo {pages} {110304} (\bibinfo {year} {2020})}\BibitemShut
  {NoStop}%
\bibitem [{\citenamefont {Peterson}\ \emph {et~al.}(2020)\citenamefont
  {Peterson}, \citenamefont {Crooks},\ and\ \citenamefont {Smith}}]{cooks}%
  \BibitemOpen
  \bibfield  {author} {\bibinfo {author} {\bibfnamefont {E.~C.}\ \bibnamefont
  {Peterson}}, \bibinfo {author} {\bibfnamefont {G.~E.}\ \bibnamefont
  {Crooks}},\ and\ \bibinfo {author} {\bibfnamefont {R.~S.}\ \bibnamefont
  {Smith}},\ }\href@noop {} {\bibfield  {journal} {\bibinfo  {journal}
  {{Quantum}}\ }\textbf {\bibinfo {volume} {4}},\ \bibinfo {pages} {247}
  (\bibinfo {year} {2020})}\BibitemShut {NoStop}%
\bibitem [{\citenamefont {Schuch}\ and\ \citenamefont
  {Siewert}(2003)}]{regensburg}%
  \BibitemOpen
  \bibfield  {author} {\bibinfo {author} {\bibfnamefont {N.}~\bibnamefont
  {Schuch}}\ and\ \bibinfo {author} {\bibfnamefont {J.}~\bibnamefont
  {Siewert}},\ }\href@noop {} {\bibfield  {journal} {\bibinfo  {journal} {Phys.
  Rev. A}\ }\textbf {\bibinfo {volume} {67}},\ \bibinfo {pages} {032301}
  (\bibinfo {year} {2003})}\BibitemShut {NoStop}%
\bibitem [{\citenamefont {Cohen}\ \emph {et~al.}(2015)\citenamefont {Cohen},
  \citenamefont {Weidt}, \citenamefont {Hensinger},\ and\ \citenamefont
  {Retzker}}]{MS}%
  \BibitemOpen
  \bibfield  {author} {\bibinfo {author} {\bibfnamefont {I.}~\bibnamefont
  {Cohen}}, \bibinfo {author} {\bibfnamefont {S.}~\bibnamefont {Weidt}},
  \bibinfo {author} {\bibfnamefont {W.~K.}\ \bibnamefont {Hensinger}},\ and\
  \bibinfo {author} {\bibfnamefont {A.}~\bibnamefont {Retzker}},\ }\href@noop
  {} {\bibfield  {journal} {\bibinfo  {journal} {New Journal of Physics}\
  }\textbf {\bibinfo {volume} {17}},\ \bibinfo {pages} {043008} (\bibinfo
  {year} {2015})}\BibitemShut {NoStop}%
\bibitem [{\citenamefont {Wang}\ \emph {et~al.}(2011)\citenamefont {Wang},
  \citenamefont {Fowler},\ and\ \citenamefont {Hollenberg}}]{threshold_2}%
  \BibitemOpen
  \bibfield  {author} {\bibinfo {author} {\bibfnamefont {D.~S.}\ \bibnamefont
  {Wang}}, \bibinfo {author} {\bibfnamefont {A.~G.}\ \bibnamefont {Fowler}},\
  and\ \bibinfo {author} {\bibfnamefont {L.~C.~L.}\ \bibnamefont
  {Hollenberg}},\ }\href@noop {} {\bibfield  {journal} {\bibinfo  {journal}
  {Phys. Rev. A}\ }\textbf {\bibinfo {volume} {83}},\ \bibinfo {pages} {020302}
  (\bibinfo {year} {2011})}\BibitemShut {NoStop}%
\bibitem [{\citenamefont {Cabrera}\ and\ \citenamefont
  {Baylis}(2007)}]{CABRERA}%
  \BibitemOpen
  \bibfield  {author} {\bibinfo {author} {\bibfnamefont {R.}~\bibnamefont
  {Cabrera}}\ and\ \bibinfo {author} {\bibfnamefont {W.}~\bibnamefont
  {Baylis}},\ }\href@noop {} {\bibfield  {journal} {\bibinfo  {journal}
  {Physics Letters A}\ }\textbf {\bibinfo {volume} {368}},\ \bibinfo {pages}
  {25 } (\bibinfo {year} {2007})}\BibitemShut {NoStop}%
\bibitem [{\citenamefont {Bowdrey}\ \emph {et~al.}(2002)\citenamefont
  {Bowdrey}, \citenamefont {Oi}, \citenamefont {Short}, \citenamefont
  {Banaszek},\ and\ \citenamefont {Jones}}]{Bowdrey}%
  \BibitemOpen
  \bibfield  {author} {\bibinfo {author} {\bibfnamefont {M.~D.}\ \bibnamefont
  {Bowdrey}}, \bibinfo {author} {\bibfnamefont {D.~K.}\ \bibnamefont {Oi}},
  \bibinfo {author} {\bibfnamefont {A.}~\bibnamefont {Short}}, \bibinfo
  {author} {\bibfnamefont {K.}~\bibnamefont {Banaszek}},\ and\ \bibinfo
  {author} {\bibfnamefont {J.}~\bibnamefont {Jones}},\ }\href@noop {}
  {\bibfield  {journal} {\bibinfo  {journal} {Physics Letters A}\ }\textbf
  {\bibinfo {volume} {294}},\ \bibinfo {pages} {258 } (\bibinfo {year}
  {2002})}\BibitemShut {NoStop}%
\bibitem [{\citenamefont {Bravyi}\ \emph {et~al.}(2018)\citenamefont {Bravyi},
  \citenamefont {Englbrecht}, \citenamefont {K{\"o}nig},\ and\ \citenamefont
  {Peard}}]{Performance_QEC_coherentErrors_II}%
  \BibitemOpen
  \bibfield  {author} {\bibinfo {author} {\bibfnamefont {S.}~\bibnamefont
  {Bravyi}}, \bibinfo {author} {\bibfnamefont {M.}~\bibnamefont {Englbrecht}},
  \bibinfo {author} {\bibfnamefont {R.}~\bibnamefont {K{\"o}nig}},\ and\
  \bibinfo {author} {\bibfnamefont {N.}~\bibnamefont {Peard}},\ }\href@noop {}
  {\bibfield  {journal} {\bibinfo  {journal} {npj Quantum Information}\
  }\textbf {\bibinfo {volume} {4}},\ \bibinfo {pages} {55} (\bibinfo {year}
  {2018})}\BibitemShut {NoStop}%
\bibitem [{\citenamefont {Wallman}\ \emph {et~al.}(2015)\citenamefont
  {Wallman}, \citenamefont {Granade}, \citenamefont {Harper},\ and\
  \citenamefont {Flammia}}]{Performance_QEC_coherentErrors_III}%
  \BibitemOpen
  \bibfield  {author} {\bibinfo {author} {\bibfnamefont {J.}~\bibnamefont
  {Wallman}}, \bibinfo {author} {\bibfnamefont {C.}~\bibnamefont {Granade}},
  \bibinfo {author} {\bibfnamefont {R.}~\bibnamefont {Harper}},\ and\ \bibinfo
  {author} {\bibfnamefont {S.~T.}\ \bibnamefont {Flammia}},\ }\href@noop {}
  {\bibfield  {journal} {\bibinfo  {journal} {New Journal of Physics}\ }\textbf
  {\bibinfo {volume} {17}},\ \bibinfo {pages} {113020} (\bibinfo {year}
  {2015})}\BibitemShut {NoStop}%
\bibitem [{\citenamefont {Huang}\ \emph {et~al.}(2019)\citenamefont {Huang},
  \citenamefont {Doherty},\ and\ \citenamefont
  {Flammia}}]{Performance_QEC_coherentErrors}%
  \BibitemOpen
  \bibfield  {author} {\bibinfo {author} {\bibfnamefont {E.}~\bibnamefont
  {Huang}}, \bibinfo {author} {\bibfnamefont {A.~C.}\ \bibnamefont {Doherty}},\
  and\ \bibinfo {author} {\bibfnamefont {S.}~\bibnamefont {Flammia}},\
  }\href@noop {} {\bibfield  {journal} {\bibinfo  {journal} {Phys. Rev. A}\
  }\textbf {\bibinfo {volume} {99}},\ \bibinfo {pages} {022313} (\bibinfo
  {year} {2019})}\BibitemShut {NoStop}%
\bibitem [{\citenamefont {Fowler}\ \emph {et~al.}(2009)\citenamefont {Fowler},
  \citenamefont {Stephens},\ and\ \citenamefont {Groszkowski}}]{threshold_3}%
  \BibitemOpen
  \bibfield  {author} {\bibinfo {author} {\bibfnamefont {A.~G.}\ \bibnamefont
  {Fowler}}, \bibinfo {author} {\bibfnamefont {A.~M.}\ \bibnamefont
  {Stephens}},\ and\ \bibinfo {author} {\bibfnamefont {P.}~\bibnamefont
  {Groszkowski}},\ }\href@noop {} {\bibfield  {journal} {\bibinfo  {journal}
  {Phys. Rev. A}\ }\textbf {\bibinfo {volume} {80}},\ \bibinfo {pages} {052312}
  (\bibinfo {year} {2009})}\BibitemShut {NoStop}%
\bibitem [{\citenamefont {Fowler}\ \emph {et~al.}(2012)\citenamefont {Fowler},
  \citenamefont {Mariantoni}, \citenamefont {Martinis},\ and\ \citenamefont
  {Cleland}}]{threshold_1}%
  \BibitemOpen
  \bibfield  {author} {\bibinfo {author} {\bibfnamefont {A.~G.}\ \bibnamefont
  {Fowler}}, \bibinfo {author} {\bibfnamefont {M.}~\bibnamefont {Mariantoni}},
  \bibinfo {author} {\bibfnamefont {J.~M.}\ \bibnamefont {Martinis}},\ and\
  \bibinfo {author} {\bibfnamefont {A.~N.}\ \bibnamefont {Cleland}},\
  }\href@noop {} {\bibfield  {journal} {\bibinfo  {journal} {Phys. Rev. A}\
  }\textbf {\bibinfo {volume} {86}},\ \bibinfo {pages} {032324} (\bibinfo
  {year} {2012})}\BibitemShut {NoStop}%
\end{thebibliography}
